\documentclass[structabstract,onecolumn]{aa}  
\usepackage[comma,authoryear]{natbib}
\usepackage[pdftex,final]{graphicx}
\usepackage[english]{babel}
\usepackage[latin1]{inputenc}
\usepackage{latexsym}
\usepackage{amssymb}
\bibpunct{(}{)}{,}{a}{}{,}%
\renewcommand{\textbf}[1]{#1}
\renewcommand{\mathbf}[1]{#1}

\newcommand{\ie}{\textit{i.e.}}
\newcommand{\eg}{\textit{e.g.}}

\newcommand{\Req}{R_\mathrm{eq}}

\newcommand{\Lmax}{L_\mathrm{max}}

\newcommand{\Nr}{N_\mathrm{r}}

\newcommand{\vect}{\vec}
\newcommand{\dpart}[2]{\frac{\partial #1}{\partial #2}}

\newcommand{\grad}{\vect{\nabla}}

\newcommand{\lapl}{\Delta}

\renewcommand{\div}{\vect{\nabla} \cdot}
\renewcommand{\l}{\ell}
\renewcommand{\d}{\partial}

\newcommand{\er}{\vect{e}_r}
\newcommand{\et}{\vect{e}_{\theta}}
\newcommand{\ep}{\vect{e}_{\phi}}
\newcommand{\es}{\vect{e}_{s}}

\newcommand{\ez}{\vect{e}_z}

\newcommand{\sint}{\sin \theta}
\newcommand{\cost}{\cos \theta}
\newcommand{\cott}{\cot \theta}

\newcommand{\rz}{r_{\zeta}}
\newcommand{\rt}{r_{\theta}}
\newcommand{\rzz}{r_{\zeta\zeta}}
\newcommand{\rzt}{r_{\zeta\theta}}
\newcommand{\rtt}{r_{\theta\theta}}

\newcommand{\uz}{u^{\zeta}}
\newcommand{\ut}{u^{\theta}}

\newcommand{\uphi}{u^{\phi}}
\newcommand{\xiz}{\xi^{\zeta}}
\newcommand{\xit}{\xi^{\theta}}
\newcommand{\xip}{\xi^{\phi}}

\newcommand{\dz}{\partial_{\zeta}}
\newcommand{\dt}{\partial_{\theta}}
\newcommand{\dphi}{\partial_{\phi}}

\newcommand{\dzz}{\partial^2_{\zeta\zeta}}
\newcommand{\dzt}{\partial^2_{\zeta\theta}}
\newcommand{\dtt}{\partial^2_{\theta\theta}}
\newcommand{\dpp}{\partial^2_{\phi\phi}}
\newcommand{\cz}{c_{\zeta}}

\newcommand{\valp}{\left[\lambda + im\Omega\right]}
\newcommand{\vlp}{\left[\omega + m\Omega\right]}
\newcommand{\ds}{\partial_s}

\begin{document}

\title{Pulsation modes in rapidly rotating stellar models based on the
       Self-Consistent Field method}

\author{D. R. Reese\inst{1}
        \and
        K. B. MacGregor\inst{2}
        \and
        S. Jackson\inst{2}
        \and
        A. Skumanich\inst{2}
        \and
        T. S. Metcalfe\inst{2}
       }

\institute{Department of Applied Mathematics,
           University of Sheffield,
           Hicks Building,
           Hounsfield Road,
           S3 7RH,
           Sheffield, UK \\
              \email{d.reese@sheffield.ac.uk}
          \and
           High Altitude Observatory, National Center for Atmospheric Research,
           Boulder, CO 80307, USA
          }

\date{}

\abstract
{New observational means such as the space missions CoRoT and Kepler and
ground-based networks are and will be collecting stellar pulsation data with
unprecedented accuracy.  A significant fraction of the stars in which pulsations
are observed are rotating rapidly.}
{Our aim is to characterise pulsation modes in rapidly rotating stellar models
so as to be able to interpret asteroseismic data from such stars.}
{The pulsation code developed in Ligni{\`e}res et al.~(2006) and Reese et
al.~(2006) is applied to stellar models based on the self-consistent field (SCF)
method (Jackson et al. 2004, 2005, MacGregor et al. 2007).}
{Pulsation modes in SCF models follow a similar behaviour to those in uniformly
rotating polytropic models, provided that the rotation profile is not too
differential.  Pulsation modes fall into different categories, the three main
ones being island, chaotic, and whispering gallery modes, which are rotating
counterparts to modes with low, medium, and high $\l-|m|$ values, respectively.
The frequencies of the island modes follow an asymptotic pattern quite similar
to what was found for polytropic models.  Extending this asymptotic formula to
higher azimuthal orders reveals more subtle behaviour as a function of $m$ and
provides a first estimate of the average advection of pulsation modes by
rotation.  Further calculations based on a variational principle confirm this
estimate and provide rotation kernels that could be used in inversion methods. 
When the rotation profile becomes highly differential, it becomes more and more
difficult to find island and whispering gallery modes at low azimuthal orders. 
At high azimuthal orders, whispering gallery modes, and in some cases island
modes, reappear.}
{The asymptotic formula found for frequencies of island modes can potentially
serve as the basis of a mode identification scheme in rapidly rotating stars
when the rotation profile is not too differential.}


\maketitle

\section{Introduction}

New observational means are and will be collecting stellar pulsation data with
unprecedented accuracy.  The space mission CoRoT has considerably lowered the
detection threshold for pulsation modes, thus allowing photometric observation
of solar-like pulsations in stars other than the Sun and increasing the
number of detected modes in early-type stars.  The forthcoming space mission
Kepler will add a wealth of pulsation data by observing a large number of stars
for a period of four years.  Other projects include the space mission PLATO as
well as ground-based networks such as SONG.

Stellar pulsations yield valuable information on the internal structure of stars
which can be used to constrain stellar evolution models.  Although a great deal
of success has been achieved in probing the internal structure of the
Sun and of a number of other stars, a number of difficulties arise for
rapidly rotating stars.  Indeed, rapid stellar rotation introduces a number of
phenomena which considerably complicate their modelling and the study of their
pulsation modes. These include centrifugal deformation, gravity darkening,
baroclinic flows and various forms of turbulence and transport phenomena
\citep[\eg\ ][] {Rieutord2006}.  As a result, the internal structure of these
stars remain difficult to probe.

Traditionally, the effects of rotation on pulsation modes have been
modelled using the perturbative approach.  In this approach, rotation is taken
into account through corrections which are added to the non-rotating solutions.
The underlying assumption in this method is that the rotation rate, $\Omega$,
can be treated as a small parameter, thus enabling one to develop the
perturbative corrections as a power series in $\Omega$.  Such series can be
extended to first \citep{Cowling1949, Ledoux1951}, second \citep{Saio1981,
Gough1990, Dziembowski1992} or third order in $\Omega$ \citep{Soufi1998,
Karami2005}.  A natural question to ask is up to what rotation rate is this
approach valid.  This remained an open question until \citet{Reese2006} applied
a non-perturbative two-dimensional approach to calculating acoustic pulsations
in polytropic stellar models and compared the results with perturbative
calculations.  Their results showed that perturbative methods remain valid only
for values which are lower than the rotation rate of many early-type stars. 
Further comparisons between the two approaches include those by
\citet{Lovekin2008}, in which more realistic models were used but at a lower
accuracy, and \citet{Ouazzani2009}, in which the effects of avoided crossings
are included in the perturbative calculations.

Due to the limitations of the perturbative method, a number of recent
studies have focused on modelling the effects of rapid rotation on stellar
acoustic pulsations using a two-dimensional approach.  \citet{Espinosa2004}
studied the effects of rapid rotation on frequency multiplets in models with a
uniform density and also briefly discussed pulsations of realistic models.  They
showed how rotation leads to highly non-uniform multiplets and causes the
frequencies of adjacent modes to pair up, thus providing a tentative explanation
for observed close frequency pairs \citep{Breger2006}. 
\citet{Lignieres2001} studied pulsation modes in a uniform density
spheroid using a perturbative method and two different numerical approaches. 
This was done in order to validate their two-dimensional numerical method before
applying it to more realistic models.  Their work was followed by
\citet{Lignieres2006, Reese2006} and \citet{Reese2008a} who did the first
accurate calculations of p-modes in rapidly rotating polytropic models.  They
investigated the limits of the perturbative approach, studied disk averaging
factors which intervene in mode visibility, compared the effects of the
centrifugal and Coriolis forces and found an empirical formula which
characterises the structure of the frequency spectrum for low degree modes.  At
the same time, \citet{Lignieres2008} and \citet{Lignieres2009} applied ray
dynamics to the study of acoustic modes in rotating polytropic models.  They
classified modes into several categories, the main ones being island,
chaotic, and whispering gallery modes which are rotating counterparts to
modes with low, intermediate, and high $\l-|m|$ values, where $\l$ is the
harmonic degree and $m$ the azimuthal order.  They showed that each category
has its own frequency organisation and provided an explanation
involving travel time integrals for the empirical formula found in
\citet{Lignieres2006} and \citet{Reese2008a}.  Finally, \citet{Lovekin2008} and
\citet{Lovekin2009} studied p-modes with low radial orders in realistic models
from \citet{Deupree1990} and \citet{Deupree1995} with both uniform and
differential rotation. They investigated how frequencies and the large and small
separations vary with uniform or differential rotation and compared their
calculations with a perturbative approach.

Before being able to interpret pulsation modes in observed stars, more progress
is needed in understanding the effects of rotation on pulsation modes.  Indeed,
although a number of important results have been established for p-modes in
polytropic models, these need to be extended to more realistic models.  The
calculations involving more realistic models have currently been limited to
small mode sets and the analysis has not been pushed far enough to see whether
similar results apply.  In what follows, we calculate pulsation modes, using the
numerical method developed in \citet{Lignieres2006} and \citet{Reese2006}, in
realistic models of rapidly rotating stars based on the Self-Consistent Field
(SCF) method \citep{Jackson2005, MacGregor2007}.  In particular, we investigate
whether a similar mode classification exists in these models, whether a similar
empirical formula applies to frequencies of modes with low $\l-|m|$ values, and
quantify the effects of using a differential profile.  The next section deals
with the SCF method and the models it produces.  The following section explains
the pulsation equations, the numerical method used for calculating the pulsation
modes and a number of tests to validate the method.  Afterwards,
Sections~\ref{sect:mildly_differential} and~\ref{sect:strongly_differential} 
describe the results for models with mildly and strongly differential rotation,
respectively.  Our conclusions are summarised in Section~\ref{sect:conclusion}.

\section{Stellar models based on the SCF method}

The SCF method is an iterative procedure for solving the equations
that govern the structure of a conservatively rotating star. The
basic approach underlying the method is to alternately solve Poisson's
equation to derive the 2D shapes of equipotential surfaces, and the
equations of mass, momentum, and energy conservation to obtain the
1D profiles of thermodynamic quantities along a radius in the rotational
equatorial plane. As described in detail in \citet{Jackson2005}, this
procedure yields a sequence of models which, under most circumstances,
converges to a model that satisfies all the equations for a prescribed
internal rotation law.

The method was first developed and used 40 years ago to compute uniformly and
differentially rotating polytropic stellar models \citep{Ostriker1968}. 
Although subsequently extended through the incorporation of more realistic input
physics \citep{Jackson1970}, application of the method was limited to massive
stars, a consequence of convergence difficulties encountered in lower mass
models with sufficiently high values of the central mass concentration
\citep[see, \eg,][]{Clement1978}.  This problem was addressed and remedied
through a reformulation of the method in which the normalised distributions of
thermodynamic quantities and the central values of those quantities are adjusted
in separate iterative loops.  The new SCF method has been implemented in
a code that utilises up-to-date input physics.  The opacities are obtained from
the tables of OPAL opacities computed by \citet{Rogers1992} and from tables of
low-temperature opacities compiled by \citet{Alexander1994}, using interpolation
subroutines written by \citet{VandenBerg1983}. The equation of state for the
stellar material is calculated according to the formula of \citet{Eggleton1973},
and the nuclear energy generation rates for hydrogen burning are from
\citet{Caughlin1988}, with the treatment of electron screening effects from
\citet{Graboske1973} for the case of equilibrium abundances of CNO isotopes. 
Energy transport in sub-photospheric convective envelopes is treated using a
standard mixing-length model  \citep[see, \eg,][]{Kippenhahn1967}, in which the
local gravitational acceleration $\vect{g}$ is replaced by the value of
$\vect{g}$ as reduced by the local centrifugal acceleration, averaged over
equipotential surfaces.  For the models utilised in the pulsation mode
computations described in subsequent sections, a value of 1.9 was adopted for
the ratio of the mixing length to the pressure scale height.  The method and the
code are both robust and rapidly convergent, and have been thoroughly tested and
validated through such applications as the interpretation of interferometric
observations of rapid rotators like the Be star Achernar \citep[][and references
therein]{Jackson2004} and an examination of the effects of differential rotation
on the structure of  stars less massive than 2  $M_{\odot}$
\citep{MacGregor2007}.

Models computed using the SCF method are chemically homogeneous, ZAMS 
models with the following abundance fractions by weight of H, He, and heavy
elements: $X=0.7112$, $Y=0.27$, and $Z=0.0188$. The rotation profile is imposed
beforehand and is conservative, \ie\ the centrifugal force derives from a
potential.  As a result, the stellar structure is barotropic -- different
thermodynamic quantities remain constant along lines of constant total
(centrifugal plus gravitational) potential.  The rotation profile used in the
present calculations was:
\begin{equation}
  \Omega(s) = \frac{\eta \Omega_{\mathrm{cr}}}
  {1+\left(\frac{\alpha s}{\Req}\right)2}
\label{eq:rotation}
\end{equation}
where $s$ is the distance from the rotation axis, $\Req$ the equatorial radius,
and $\Omega_{cr}$ the break-up rotation rate at $\Req$.  The
parameters $\eta$ and $\alpha$ determine how rapid and differential
the rotation is.  In particular, the ratio between the polar and equatorial
rotation rate is $1+\alpha^2$.  The associated angular momentum increases with
$s$ thereby satisfying the dynamical part of the Solberg-H{\o}iland criterion
for stability.  Various forms of shear instability may, however, be present if
the rotation profile becomes too differential, \ie\ if $\alpha$ becomes too
large \citep[\eg\ ][]{Zahn1974}.  Also, as explained in \citet{Zahn1993} and
\citet{Rieutord2006c}, baroclinic flows occur in radiative zones of rapidly
rotating stars thus leading to a non-conservative rotation profile.  Exploring
these effects is, however, beyond the scope of this paper.

Equation~(\ref{eq:rotation}) corresponds to a rotation profile in which the
rotation rate decreases with $s$.  Such profiles can be used to construct highly
distorted configurations.  Indeed, the stellar core can be made to rotate quite
rapidly since the local break-up velocity is larger than at the equator.  This
type of model was used to try to explain Achernar's extreme oblateness
\citep{Jackson2004}.  The SCF method can also produce models with a rotation
rate that increases with distance from the rotation axis.  This resembles
somewhat the solar rotation profile in which the rotation rate increases with
decreasing latitude in the convection zone \citep{Schou1998, Thompson2003}.

\subsection{The pulsation equations}

In order to derive the set of equations which govern acoustic pulsation modes 
in a differentially rotating star, we start by representing the differential
rotation by a permanent background flow $\vect{v}_o = s \Omega(s) \ep$. In what
follows, we will work with cylindrical coordinates $(s,z,\phi)$ and their
associated unit vectors $(\es,\ez,\ep)$.  We write out the Eulerian perturbation
to various equations, starting with Euler's equation, and only keep first order
linear terms:
\begin{equation}
\rho_o \dpart{\vect{v}}{t} + \rho \vect{v}_o \cdot \grad \vect{v}_o +
\rho_o \vect{v} \cdot \grad \vect{v}_o + \rho_o \vect{v}_o \cdot \grad \vect{v}
= - \grad p + \rho \vect{g}_\mathrm{o} - \rho_o \grad \Psi,
\label{eq:Euler1}
\end{equation}
where quantities with the subscript ``$o$'' are equilibrium quantities, and
those without a subscript Eulerian perturbations.  The quantity
$\vect{g}_\mathrm{o}$ is the background gravity excluding
the centrifugal acceleration.  The different terms on the left hand side
of Eq.~(\ref{eq:Euler1}) can be worked out explicitly in terms of $\Omega(s)$:
\begin{eqnarray}
\vect{v}_o \cdot \grad \vect{v}_o &=& -s \Omega^2 \es, \\
\vect{v} \cdot \grad \vect{v}_o   &=& \vect{\Omega} \times \vect{v} + v_s s \d_s \Omega \ep, \\
\vect{v}_o \cdot \grad \vect{v}   &=& \vect{\Omega} \times \vect{v} + im\Omega\vect{v},
\end{eqnarray}
where we have assumed an $e^{im\phi}$ azimuthal dependence for $\vect{v}$ and
$\vect{\Omega} = \Omega\ez$.  The first term corresponds to the centrifugal
acceleration and the sum of the next two includes the Coriolis force.  Combining
these equations with Eq.~(\ref{eq:Euler1}) yields:
\begin{equation}
\valp \rho_o \vect{v}  =  - \grad p + \rho \vect{g}_\mathrm{eff}
                          - \rho_o \grad \Psi
                          - 2 \vect{\Omega} \times \rho_o\vect{v}
                          - \rho_o s\dpart{\Omega}{s} v_s \ep.
\label{eq:Euler2}
\end{equation}
where we have assumed an $e^{\lambda t}$ time dependence for $\vect{v}$ and
where $\vect{g}_\mathrm{eff} = \grad p_o/\rho_o$ is the background effective
gravity which includes the centrifugal acceleration.  The Eulerian perturbation
to the continuity equation gives:
\begin{equation}
\dpart{\rho}{t} + \div \left( \rho_o \vect{v} \right) + \div \left(
\rho \vect{v}_o \right) = 0.
\label{eq:continuity1}
\end{equation}
In terms of $\Omega$, this becomes:
\begin{equation}
\valp \rho = -\vect{v} \cdot \grad \rho_o - \rho_o \div \vect{v}.
\label{eq:continuity2}
\end{equation}
The Eulerian perturbation to Poisson's equation is simply:
\begin{equation}
\lapl \Psi = 4\pi G \rho,
\label{eq:Poisson1}
\end{equation}
where $G$ is the gravitational constant.  These equations are then supplemented
by the adiabatic relation between the pressure and density perturbations
which takes on the following form:
\begin{equation}
\dpart{p}{t} + \vect{v} \cdot \grad p_o + \vect{v}_o \cdot \grad p
= c_o^2
\left[
\dpart{\rho}{t} + \vect{v} \cdot \grad \rho_o + \vect{v}_o \cdot \grad \rho
\right]
\label{eq:energy1}
\end{equation}
where $c_o^2 = \frac{\Gamma_1 p_o}{\rho_o}$ is the square of the sound velocity
and $\Gamma_1$ the adiabatic exponent.  Provided that $\vect{v}_o \cdot \grad
p_o = \vect{v}_o \cdot \grad \rho_o = \vect{v}_o \cdot \grad c_o^2 = 0$, this
form can be shown to be equivalent to $\delta p = c_o^2 \delta \rho$ where
$\delta p$ and $\delta \rho$ are the Lagrangian pressure and density
perturbations, respectively.  This leads to the following equation after some
manipulations:
\begin{equation}
\valp \left(p - c_o^2 \rho \right) =
              \left[ -\grad p_o + c_o^2 \grad \rho_o\right] \cdot \vect{v} \\
\label{eq:energy2}
\end{equation}

\subsection{Non-dimensional form}
These equations are then put into non-dimensional form using the following 
length, density and pressure scale factors:
\begin{equation}
\Req, \qquad p_c, \qquad \rho_c,
\end{equation}
where the subscript ``$c$'' refers to the centre of the star, and $\Req$ is the
equatorial radius. This gives a time scale $t_\mathrm{ref}$ defined as:
\begin{equation}
t_\mathrm{ref} = \left( \frac{\rho_c \Req^2}{p_c} \right)^{1/2}
\end{equation}

Based on these scale factors, all of the above equations remain the same as in
dimensional form, except for Poisson's equation where a non-dimensional factor
$\Lambda = 4\pi G \rho_c^2 \Req^2/p_c$ appears:
\begin{equation}
\lapl \Psi = \Lambda \rho.
\label{eq:Poisson2}
\end{equation}

\subsection{Spheroidal geometry}
\label{sect:geometry}
In order to achieve higher accuracy when solving these equations numerically, a
coordinate system which follows the shape of the star is introduced.  This new
coordinate system $(\zeta,\theta,\phi)$ can be related to the usual
spherical coordinate system $(r,\theta,\phi)$ via the following relationship:
\begin{equation}
r(\zeta,\theta) = (1-\varepsilon)\zeta+\frac{5\zeta^3-3\zeta^5}{2}
                 \left( R_s(\theta) - 1 + \varepsilon \right),
\label{eq:domain1}
\end{equation}
for $\zeta \in [0,1]$.  When $\zeta = 1$, $r$ coincides with the stellar
surface, $R_s(\theta)$.  The variables $\theta$ and $\phi$ remain the same in
both systems.  A second domain is added around the first, in which $r$ is given
by:
\begin{equation}
r(\zeta,\theta) = 2\varepsilon + (1-\varepsilon) \zeta
                 + \left( 2\zeta^3 - 9\zeta^2+12\zeta-4\right)
                   \left( R_s(\theta) - 1 - \varepsilon \right),
\label{eq:domain2}
\end{equation}
for $\zeta \in [1,2]$.  With these definitions, $r$ and $\rz \equiv
\partial_\zeta r$ remain continuous across $\zeta = 1$.  As $\zeta$
approaches $0$ or $2$, this coordinate system behaves like a spherical
coordinate system: the constant $\zeta$-lines become spherical and $\rz$ becomes
independent of $\theta$.  This is important as it simplifies the regularity
conditions in the centre and the boundary condition on the perturbation to the
gravity potential on the outer boundary.  The same coordinate system was used
in \citet{Lignieres2006} and \citet{Reese2006} and is based on
\citet{Bonazzola1998}.  The stellar model is then interpolated onto a grid based
on this new coordinate system.

Another possibility would be to base the radial coordinate on the
equipotentials.  This has the advantage of simplifying the pulsation equations
because terms such as $\partial_\theta \rho_o$ and $\partial_\theta p_o$ vanish.
However, this requires using numerical rather than analytical differentiation
when calculating terms with radial derivatives such as $\rz$, thereby reducing
the accuracy of the results.  Furthermore, the regularity conditions
in the centre of the star become more complicated as the equipotentials
do not in general become circular towards the centre.

Based on the coordinate system presented above, the continuity
equation becomes:
\begin{equation}
\label{eq:spheroidal.continuity}
\valp \rho  =  
              - \frac{\zeta^2 \dz \rho_o}{r^2 \rz} \uz
              - \frac{\zeta \dt \rho_o}{r^2 \rz} \ut
              - \frac{\zeta^2 \rho_o}{r^2 \rz}\left[
                \frac{\dz \left( \zeta^2 \uz \right)}{\zeta^2}
              + \frac{\dt \left( \sint \ut \right)}{\zeta \sint}
              + \frac{\dphi \uphi}{\zeta \sint}
               \right], \\
\end{equation}
Euler's equation takes on the following form:
\begin{eqnarray}
\label{eq:spheroidal.Euler1}
\valp \rho_o \left[ \frac{\zeta^2 \rz \uz}{r^2} 
        +\frac{\zeta \rt \ut }{r^2}\right] &=&
         \rho_o \frac{2 \Omega \zeta \sint \uphi}{r}  
        - \dz p + \frac{\dz p_o}{\rho_o} \rho
        - \rho_o \dz \Psi, \\
\noalign{\smallskip} 
\label{eq:spheroidal.Euler2}
\valp \rho_o \left[ \frac{\zeta^2 \rt \uz}{r^2} + 
          \frac{\zeta(r^2+\rt^2) \ut}{r^2\rz} \right] &=&
          \rho_o \frac{2 \Omega \zeta (\rt \sint + r \cost)\uphi}{r\rz} 
         -\dt p + \frac{\dt p_o}{\rho_o} \rho
         -\rho_o \dt \Psi, \\
\noalign{\smallskip} 
\label{eq:spheroidal.Euler3}
\valp \rho_o \frac{\zeta\uphi}{\rz} &=&
           -\rho_o \frac{2 \Omega \zeta^2 \sint\uz}{r}
           -\rho_o \frac{2 \Omega \zeta(\rt \sint + r \cost)\ut}{r\rz}
           -\frac{\dphi p}{\sint} \nonumber \\
       & & -\rho_o \frac{\dphi \Psi}{\sint}
           -\rho_o \sint \left( \ds \Omega \right) 
            \left[ \zeta^2 \sint \uz 
          + \frac{\zeta(\rt \sint + r \cost)}{\rz} \ut \right],
\end{eqnarray}
the adiabatic relation is given by:
\begin{equation}
\label{eq:spheroidal.energy}
\valp \left( p - c_o^2 \rho \right)  =
      \frac{\zeta^2}{r^2 \rz} 
      \left( - \dz p_o + c_o^2 \dz \rho_o \right) \uz
    + \frac{\zeta}{r^2 \rz} 
      \left( - \dt p_o + c_o^2 \dt \rho_o \right) \ut, \\
\end{equation}
and Poisson's equation takes on the following form:
\begin{equation}
\label{eq:spheroidal.Poisson}
0  =  \frac{r^2 + \rt^2}{r^2 \rz^2}  \dzz \Psi
+\cz  \dz \Psi 
-\frac{2\rt}{r^2 \rz} \dzt \Psi
+\frac{1}{r^2} \lapl_{\theta \phi} \Psi
- \Lambda \rho,
\end{equation}
where $\uz$, $\ut$, and $\uphi$ are the three velocity components (see
below for details) and terms of the form $\rz$, $\rt$ ... are different
derivatives of $r$ based on Eqs.~(\ref{eq:domain1}) and~(\ref{eq:domain2}). 
Explicit expression for $\cz$ and $\lapl_{\theta\phi}$ are as follows:
\begin{eqnarray}
\cz &=& \frac{1}{r^2 \rz^3} \left( 2 \rz \rt \rzt - r^2 \rzz - \rz^2 \rtt 
+ 2 r\rz^2 - \rt^2 \rzz -  \rz^2 \rt \cott \right), \\
\lapl_{\theta\phi} &=& \dtt + \cott \dt + \frac{1}{\sin^2 \theta}\dpp.
\end{eqnarray}

The above expressions are obtained by using tensorial expressions for
the differential operators which intervene and working them out explicitly. 
Furthermore, the components to the velocity are written on the following basis:
\begin{equation}
\begin{array}{lllll}
\vect{a}_{\zeta}  &=& \displaystyle \frac{\zeta^2}{r^2 \rz} \vect{E}_{\zeta}
                  &=& \displaystyle \frac{\zeta^2}{r^2} \er, \\
\vect{a}_{\theta} &=& \displaystyle \frac{\zeta}{r^2 \rz} \vect{E}_{\theta}
                  &=& \displaystyle \frac{\zeta}{r^2 \rz} \left( \rt \er + r \et
                      \right), \\
\vect{a}_{\phi}   &=& \displaystyle \frac{\zeta}{r^2 \rz \sint} \vect{E}_{\phi}
                  &=& \displaystyle \frac{\zeta}{r \rz} \ep,
\end{array}
\end{equation}
where $\left\{\vect{E}_i\right\}$ and $\left\{\vect{e}_i\right\}$ are
the natural and spherical basis, respectively.  When the star becomes spherical,
the basis $\left\{\vect{a}_i\right\}$ converges to $\left\{\vect{e}_i\right\}$. 
Apart from some multiplicative factors, Euler's equation is expressed on the
dual basis, as is done in \citet{Reese2006}, since this has the advantage of
limiting the effective gravity to the radial component at the stellar surface. 
More details on how to derive the above expressions can be found in
\citet{Reese2006PhD} and references therein.

Equations~(\ref{eq:spheroidal.continuity})-(\ref{eq:spheroidal.Poisson})
were completed with a number of boundary conditions which ensure that the
solution remains regular in the centre, the Lagrangian pressure perturbation
vanishes on the stellar surface and the perturbation to the gravitational
potential goes to zero at an infinite distance from the star.

\subsection{Numerics}

The above equations were then projected onto the spherical harmonic basis in the
same way as was done in \citet{Lignieres2006} and \citet{Reese2006}.  This is
achieved by expressing the different unknowns as a sum of scalar or vectorial
spherical harmonics multiplied by unknown radial functions, and then projecting
the equations themselves onto the spherical harmonic basis, using
Gaussian quadrature to numerically perform the integrations.  The resultant
system is an infinite system of coupled ordinary differential equations in terms
of the radial variable $\zeta$ which is truncated at a maximal harmonic degree
$\Lmax$.  The solution to this system yields the radial functions used
in the harmonic decomposition of the different unknowns.

This system of ordinary differential equations is discretised using one of three
methods: a spectral method based on Chebyshev polynomials, finite differences or
a polynomial spline-based method.  In the latter two cases, the order
can be adjusted.  When applying these methods, the stellar model is
interpolated onto either a higher resolution uniform grid or a Chebyshev
Gauss-Lobatto collocation grid using cubic spline interpolation.  For the
spectral method, this is analogous to what was done in \citet{Dintrans2000} in
which a $1.5\,\,M_\odot$ CESAM model was interpolated onto the same type of
collocation grid before being used to calculate gravito-inertial modes. 

After discretisation, the eigenvalue problem is in algebraic form:
$\mathcal{A}v = \lambda \mathcal{B} v$ where $\mathcal{A}$ and $\mathcal{B}$ are
square matrices.  With a suitable choice of variables, these matrices can be
made real, thereby reducing the computational cost.  Also, pulsation modes are
either symmetric or anti-symmetric with respect to the equator, so that only
spherical harmonics of the same parity are needed to describe them. The problem
is then solved numerically using the Arnoldi-Chebyshev algorithm \citep[\eg\ 
][]{Chatelin1988} around different target frequencies, called frequency
shifts.  In what follows, most of the calculations have been done using a
4$^\mathrm{th}$ order finite difference approach.  The angular resolution was
typically $\Lmax = 80$ and the radial resolution $\Nr = 501$.

\subsection{Accuracy of the calculations}

\begin{figure}[h]
  \begin{center}
  \begin{tabular}{cc}
  \textbf{Intermediate radial order} & 
  \textbf{High radial order} \\
  \includegraphics[width=9cm]{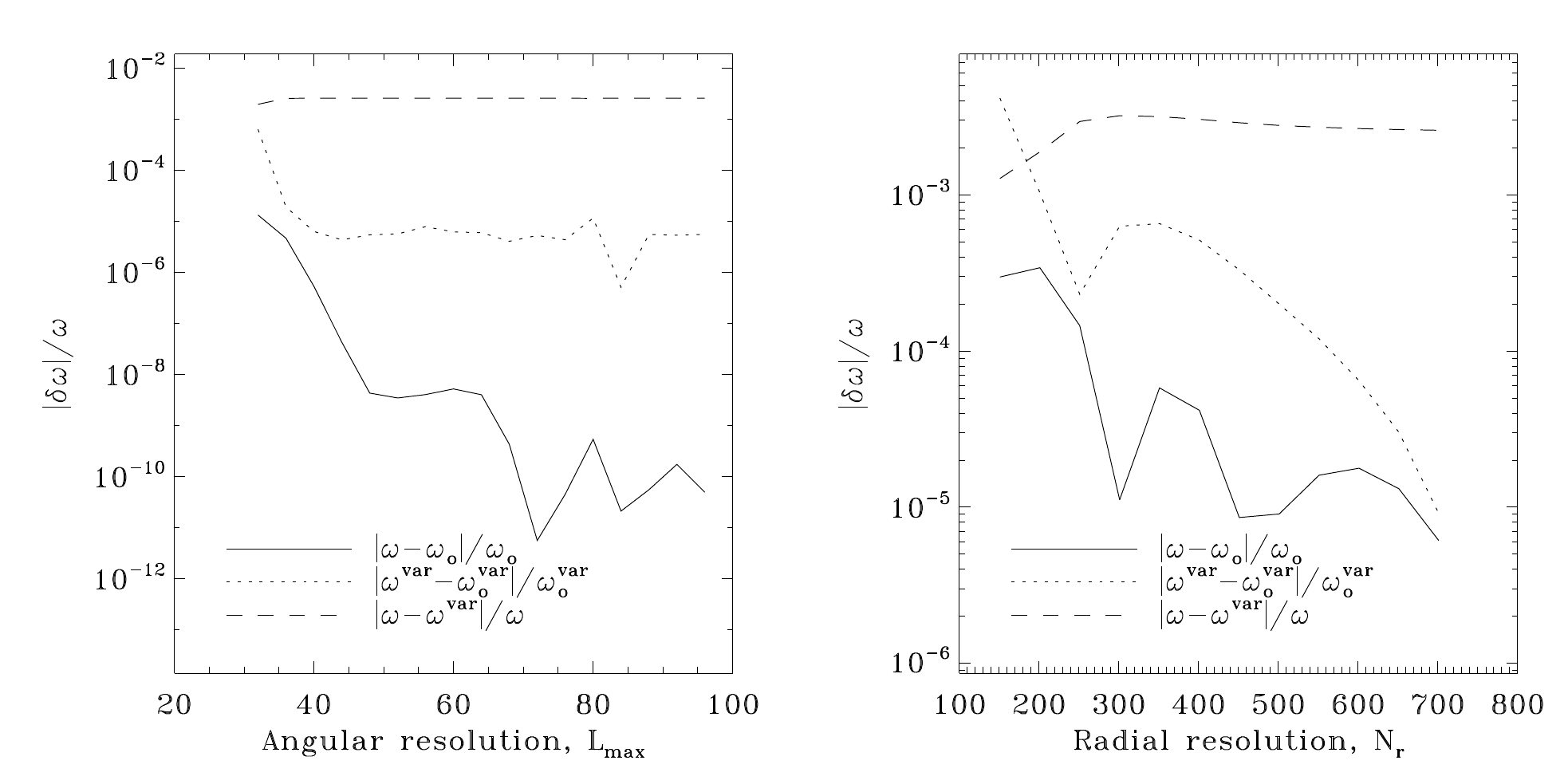} &
  \includegraphics[width=9cm]{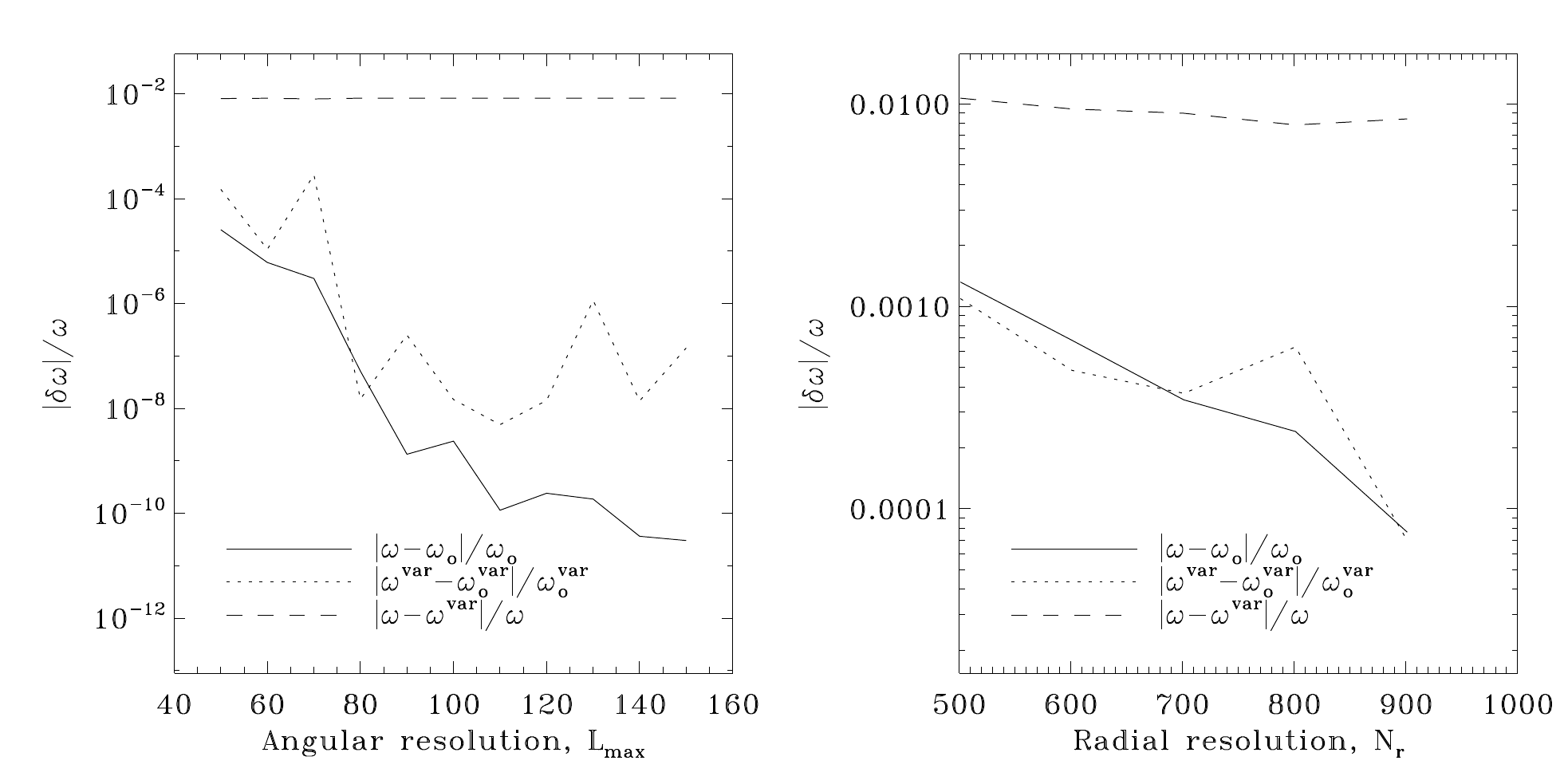}
   \end{tabular}
  \end{center}
  \caption{Evolution of various forms of the frequency error with $\Lmax$ and
           $\Nr$ for two pulsation modes in an 25 $M_{\odot}$ star rotating
           uniformly at $60\%$ of the break-up velocity.  The mode on the left
           corresponds to $\tilde{n} = 16$ and the one the right to $\tilde{n} =
           50$ (the meaning of $\tilde{n}$ is given in
           Section~\ref{sect:mildly_differential} and illustrated in
           Fig.~\ref{fig:comparison}).  The solid and dotted lines correspond to
           the relative frequency error.  This first case uses the numerical
           frequency and the second is based on the variational frequency. The
           dashed curve shows the relative difference between the numerical and
           variational frequencies.  The frequencies $\Omega_o$ and
           $\Omega_o^{\mathrm{var}}$ are the numerical and variational
           frequencies calculated at highest resolution (\ie\ $\Lmax = 100$,
           $\Nr = 751$ for the mode on the left and $\Lmax = 160$, $\Nr = 1001$
           for the mode on the right).  As can be seen in the figures, the
           numerical frequencies are very stable as a function of the
           resolutions, and the variational frequency somewhat less stable.
           Also, a discrepancy remains between the two types of frequencies.}
  \label{fig:verif}
\end{figure}

Various tests can be used to assess the accuracy of the calculations.   A first
test consists in following the evolution of the frequency error as a function of
the radial and angular resolution.  The solid lines in Fig.~\ref{fig:verif} show
the evolution of the relative error on the numerical frequency for two modes in
a $25$ $M_{\odot}$ model uniformly rotating at $60\%$ of the break-up rotation
rate, using the frequencies calculated at highest resolution as references.  The
first two panels apply to an $\tilde{n} = 16$ mode and the other two to
$\tilde{n} = 50$ (see Section~\ref{sect:mildly_differential} for the definition
of $\tilde{n}$).  As is evident from the figure, the stability of the numerical
frequencies is very good, especially for the angular resolution where spectral
convergence seems to be achieved.

The right two panels of Fig.~2 of \citet{Reese2008b} show similar curves for a
pulsation mode in a $1.8$ $M_{\odot}$ star rotating uniformly at $90\%$ of the
break-up rotation rate. In this case the results were not as good.  As explained
in \citet{Reese2008b}, evaluating the error in this case was not entirely
straightforward due to difficulties in identifying the correct mode at different
resolutions.  Indeed, at such high rotation rates, regular modes interact much
more with chaotic ones thus distorting their geometric features. Furthermore,
the amount of interaction between the different modes seems to depend on the
numerical resolution.

Although the problem is expressed in terms of real matrices and
frequencies are searched for around real target frequencies, complex conjugate
solutions sometimes appear.  For instance, two of the calculations in the panel
to the far right of Fig.~\ref{fig:verif} (at $\Nr=601$ and $\Nr=801$) correspond
to complex solutions.  The imaginary parts are most likely due to numerical
inaccuracies as these solutions are replaced by real solutions at other
numerical resolutions.  Their relative magnitude ($3\times 10^{-5} - 10^{-4}$)
suggests a comparable accuracy on the corresponding frequencies.

Another test consists in applying a variational formula on the eigenmodes to
yield an independent value for the frequency.  According to the variational
principle, the error on the ``variational frequency'' is proportional to the
square of the error on the eigenmode, thus minimising its effect provided it is
sufficiently small \citep{Christensen-Dalsgaard1994}.  By comparing this value
to the original (numerical) frequency, it is possible to estimate the accuracy
of the calculation.  In what follows, we calculated variational frequencies
using the following formula which is only valid for uniform rotation:
\begin{eqnarray}
0 &=& \left(\omega_\mathrm{var}+m\Omega\right)^2 \displaystyle \int_V \rho_o
      \|\vect{v}\|^2 dV +2 i \left(\omega_\mathrm{var}+m\Omega\right) \int_V
      \rho_o \vect{\Omega} \cdot \left( \vect{v}^* \times \vect{v} \right) dV -
      \int_V \rho_o N_o^2 \left| \vect{v} \cdot \vect{e}_g \right|^2 dV
      \nonumber \\
  & & - |\omega+m\Omega|^2 \left( \int_V \frac{|p|^2 dV}{\rho_o c_o^2} -
      \frac{1}{\Lambda}\int_{V_{\infty}} \| \grad \Psi \|^2 dV \right),
\label{eq:variational}
\end{eqnarray}
where $\omega$ is the numerical frequency, $\omega_{\mathrm{var}}$ the
variational frequency, $m$ the azimuthal order, $V$ the volume of the
star, $V_{\infty}$ infinite space, and $\vect{e}_g$ the unit vector in the same
direction as the effective gravity.  The geometric term $m\Omega$ comes
from the fact that the pulsation frequencies are expressed in an inertial frame.
In Section \ref{sect:omega_eff}, we give a more general variational formula
which is also valid for differential rotation, but is expressed in terms of the
Lagrangian displacement rather than the Eulerian velocity perturbation. Such a
formulation gives comparable results as Eq.~\ref{eq:variational}, \textit{i.e.}
$\delta \omega/\omega \lesssim 10^{-3}-10^{-2}$, even for the most differential
rotation profiles.

The dashed lines in Fig.~\ref{fig:verif} show the relative difference between
the numerical and variational frequencies.  As can be seen in the figure, these
differences are much larger than the variations caused by modifying the
resolution.  A third set of curves, the dotted lines, show the relative
error on the variational frequencies when using the variational frequency at
highest resolution as a reference.  From these, we deduce that the variational
frequencies do converge to a specific value, but at a slower rate than the
numerical frequencies, which is the opposite of what we would expect from the
variational principle.  Furthermore, the limit of the variational frequencies is
different than that of the numerical frequencies, as can be seen from the dashed
curves.  Such a discrepancy can occur if the models are not in perfect
hydrostatic equilibrium due typically to numerical inaccuracies.  Indeed,
hydrostatic equilibrium is implicitly assumed when deriving the variational
formula, and deviations from this state will tend to produce errors which are
independent of the resolution of the eigenfunctions.  Nonetheless, these
discrepancies remain small when the rotation rate is not too close to break-up
and probably affect the different modes in a similar way for a given model so
that the analysis in the rest of the article is not likely to be affected.

\begin{figure}[h]
  \begin{tabular}{ccc}
  \includegraphics[width=55mm]{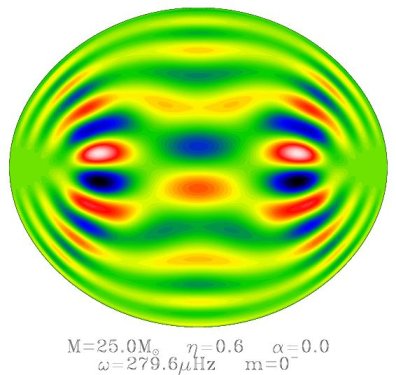}   &   
  \includegraphics[width=55mm]{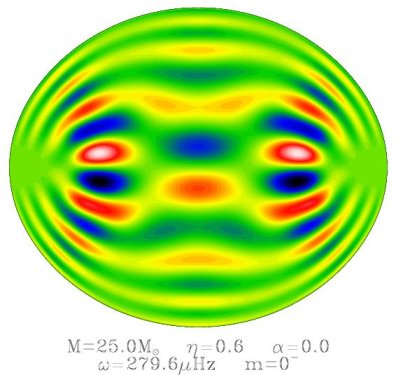} &   
  \includegraphics[width=55mm]{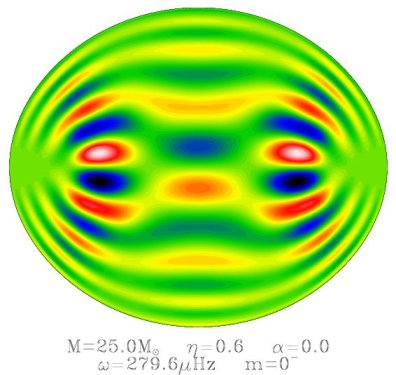} \\  
  Finite differences &
  Chebyshev polynomials &
  Polynomial splines
  \end{tabular}
  \caption{Comparison of a pulsation mode calculated using finite differences
           (\textit{left}), a spectral method (\textit{centre}), and polynomial
           splines (\textit{right}) to discretise the equations in the radial
           direction.  These and other similar plots show a meridional
           cross-section of the Eulerian pressure perturbation divided by
           the square root of the background density profile so as to
           bring out near surface regions.  The difference on the frequencies is
           less than $0.1\,\,\mu$Hz.}
  \label{fig:dertype}
\end{figure}

Finally, a last test consists in applying different numerical techniques to
calculate the eigenmodes and seeing if they give similar results. 
Figure~\ref{fig:dertype} shows such a comparison.  The mode on the left is
calculated using $4^\mathrm{th}$ order finite differences in the radial
direction, the one in the middle a spectral method based on Chebyshev
polynomials and the one on the right $4^\mathrm{th}$ order polynomial splines.
The angular resolution was very similar for the three cases and the radial
resolution went from $\Nr=101$ for the calculation based on Chebyshev
polynomials to $\Nr=301$ for the spline-based calculation.  As can be seen in
the figure, the three calculations yield very similar results, and the
corresponding frequencies are less than $0.1\,\,\mu$Hz apart. 
Furthermore, as will be explained later on, some pulsation modes were
calculated using the Lagrangian displacement rather than the Eulerian velocity
perturbation.  When comparing the two methods, differences on the frequencies
are very small for $m=0$ and can be larger for $m \neq 0$ (for example, $\delta
\omega/\omega = 10^{-3}$ for the mode represented in
Fig.~\ref{fig:high_m_high_alpha}, right panel), thereby providing yet another
verification on the pulsation mode calculations.

Overall, these tests indicate a good numerical stability both with respect to
the numerical resolution and the choice of numerical method.  The tests on the
variational principle, on the other hand, show that some numerical difficulties
remain, possibly resulting from a loss of precision on the stellar models. 
Furthermore, the accuracy is not as good when the rotation rate approaches
break-up, as shown in \citet{Reese2008b}.  Before doing accurate comparisons
with actual observations, these difficulties will need to be addressed. 
Nonetheless, these are not expected to change the basic behaviour of the
pulsation modes nor the results in following sections.

\section{Uniform or nearly uniform rotation profile}
\label{sect:mildly_differential}

\subsection{Mode classification}

\begin{figure}
  \begin{center}
  \begin{tabular}{cccc}
   & Island & Chaotic & Whispering gallery \\
   & Low $\l-|m|$ & Medium $\l-|m|$ & High $\l-|m|$ \\
  \rotatebox{90}{\hspace*{4mm}\textbf{Polytropic}} &
  \includegraphics[height=4cm]{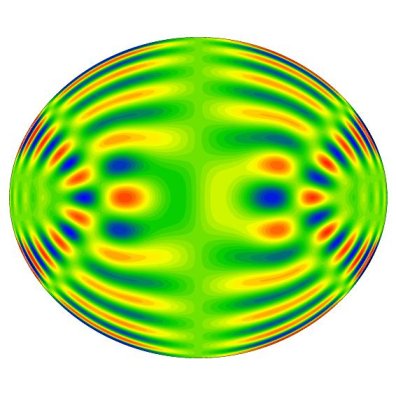} &
  \includegraphics[height=4cm]{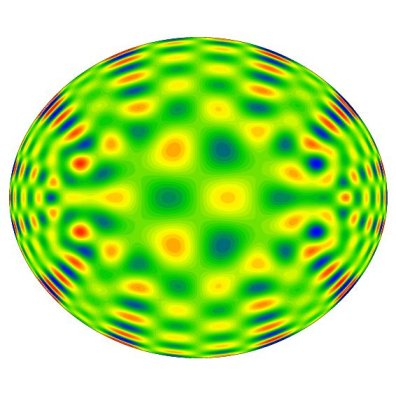} &
  \includegraphics[height=4cm]{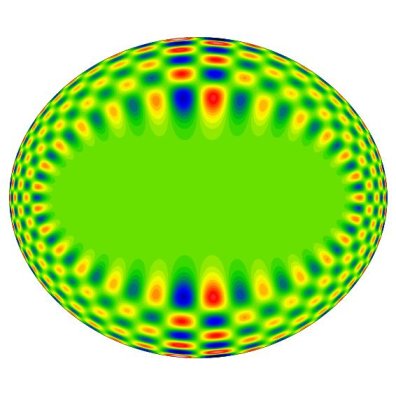} \\
  \rotatebox{90}{\hspace*{8mm}\textbf{SCF}} &
  \includegraphics[height=4cm]{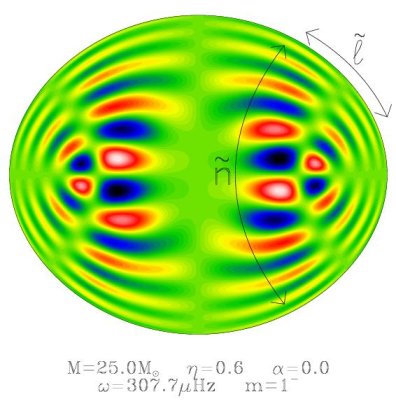} &
  \includegraphics[height=4cm]{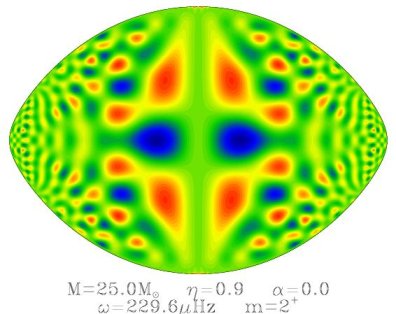} &
  \includegraphics[height=4cm]{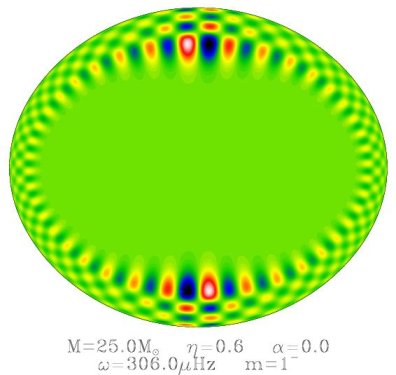}
  \end{tabular}
  \end{center}
  \caption{A comparison between pulsation modes in polytropic models and models
  based on the SCF method.  The same three categories apply in both cases as can
  be seen by the analogous geometric structure.  The quantum numbers $\tilde{n}$
  and $\tilde{\l}$, which only apply to island modes, are the number of nodes in
  the directions indicated in the lower left plot.}
  \label{fig:comparison}
\end{figure}

As was stated above, \citet{Lignieres2008} and \citet{Lignieres2009} have
previously shown that for rotating polytropic models, pulsation modes fall into
the following main categories: island, chaotic, and whispering gallery
modes.  We have found that a similar classification also applies to pulsation
modes in SCF models with uniform or mildly differential rotation (at least up to
$\alpha = 0.4$, which, based on Eq.~(\ref{eq:rotation}), gives an equatorial
rotation rate which is $84\%$ of the polar rotation rate). 
Figure~\ref{fig:comparison} compares pulsation modes from both types of models. 
As can be seen in the figure, corresponding modes with an analogous geometric
structure are also present in SCF models.

\subsection{Quantum numbers for island modes}

As was also the case for polytropic models, it is possible to introduce a new
set of quantum numbers $(\tilde{n},\,\tilde{\l},m)$ based on the geometry of
island modes (see lower left plot in Fig.~\ref{fig:comparison}).  These quantum
numbers then intervene in a new asymptotic formula which describes the frequency
organisation of these modes:
\begin{equation}
      \omega = \tilde{n} \Delta_{\tilde{n}}
             + \tilde{\l}\Delta_{\tilde{\l}}
             + m^2 \Delta_{\tilde{m}}
             - m \Omega_{\mathrm{fit}}
             + \tilde{\alpha}
\label{eq:asymptotic}
\end{equation}
where $\Delta_{\tilde{n}},\,\Delta_{\tilde{\l}},\,\Delta_{\tilde{m}}$ and
$\tilde{\alpha}$ are free parameters which depend on stellar structure.  The
parameter $\Omega_{\mathrm{fit}}$ corresponds to the rotation rate but is
treated as a free parameter.  This formula is quite similar to the one
introduced in \citep{Lignieres2008, Reese2008a} except that $|m|$ has been
replace by $m^2$ as this provides a slightly more accurate fit to the
frequencies.  The reason why $|m|$ had been obtained in \citet{Reese2008a} is
because the formula was first derived for the quantum numbers $(n,\,\l,\,m)$,
where a term proportional to $|m|$ dominates, and then adapted to
$(\tilde{n},\,\tilde{\l},\,m)$.

Table~\ref{tab:asymptotic} gives the values of these parameters for selected SCF
models as well as for a polytropic model.  The parameters were calculated from a
sparse frequency set and are therefore subject to some error. The ranges
on the quantum numbers are $10 \leq \tilde{n} \leq 26$, $0 \leq \tilde{\l} \leq 3$
and $-2 \leq m \leq 2$.  Furthermore, due to difficulties in mode identification, the
parameters given for the most rapidly rotating models $(\eta = 0.9)$ must be
taken with caution.  Nonetheless, a second calculation based on a more complete
mode set shows that these values provide a reasonable estimate for 2 of the
models (see following section).  The true value or range of values for the
rotation rate, $\Omega_{\mathrm{real}}$, is also provided and shows \textit{a
posteriori} that $\Omega_{\mathrm{fit}}$ does approximately correspond to the
rotation rate.

\begin{table}[htbp]
\begin{center}
\caption{Parameters for the asymptotic formula Eq.~(\ref{eq:asymptotic}).}
\label{tab:asymptotic}
\begin{tabular}{*{11}{c}}
\hline
\hline
$\displaystyle \frac{M}{M_{\odot}}$ & 
$\eta$ &
$\alpha$ &
$N_{\mathrm{modes}}$ &
$\!\!\!\begin{array}{c} \displaystyle \Delta_{\tilde{n}} \\ (\mu\mathrm{Hz}) \end{array}\!\!\!$ & 
$\displaystyle \frac{\Delta_{\tilde{\l}}}{\Delta_{\tilde{n}}}$ &
$\displaystyle \frac{\Delta_{\tilde{m}}}{\Delta_{\tilde{n}}}$ &
$\displaystyle \frac{\tilde{\alpha}}{\Delta_{\tilde{n}}}$ &
$\displaystyle \frac{\Omega_{\mathrm{fit}}}{\Delta_{\tilde{n}}}$  &
$\displaystyle \frac{\Omega_{\mathrm{real}}}{\Delta_{\tilde{n}}}$  &
$\displaystyle \frac{\left< \delta \omega^2 \right>^{1/2}}{\Delta_{\tilde{n}}}$ \\
\hline
poly$^{\star}$ & 0.6 & 0.0 & 84 & 36.7 & 0.66 & 0.029 & 2.92 & 0.827 & 0.838 & 0.047 \\
           1.7 & 0.7 & 0.0 & 40 & 37.1 & 0.77 & 0.018 & 3.52 & 0.975 & 0.982 & 0.023 \\
           1.8 & 0.9 & 0.0 & 11 & 33.5 & 0.42 & 0.011 & 2.86 & 1.157 & 1.167 & 0.038 \\
          25.0 & 0.6 & 0.0 & 39 & 15.2 & 0.79 & 0.016 & 3.39 & 0.947 & 0.969 & 0.030 \\
          25.0 & 0.6 & 0.2 & 31 & 15.1 & 0.85 & 0.018 & 3.63 & 0.944 & 0.947-0.987 & 0.045 \\
          25.0 & 0.6 & 0.4 & 31 & 15.5 & 0.90 & 0.050 & 3.37 & 0.915 & 0.830-0.988 & 0.059 \\
          25.0 & 0.9 & 0.0 & 24 & 12.4 & 0.70 &-0.002 & 3.41 & 1.380 & 1.387 & 0.033 \\ 
\hline
\end{tabular}
\end{center}
\begin{list}{}{}
\item[] Values of the different parameters from Eq.~(\ref{eq:asymptotic}) for
selected SCF models as well as for a polytropic model (first line).  The first
three columns identify the model, where $\eta$ and $\alpha$ come from
Eq.~(\ref{eq:rotation}).  These parameters were based on a sparse mode set (the
number of modes being indicated by $N_{\mathrm{mode}}$) and are therefore subject
to error.  The last column contains the average deviation between asymptotic
frequencies based on Eq.~(\ref{eq:asymptotic}) and the numerical frequencies.
\item[] $^\star$Polytropic model with $N = 3$, $M = 1.7 M_{\odot}$ and $\Req = 1.84
         R_{\odot}$.  These are also the mass and equatorial radius of the model
         on the next line.
\end{list}
\end{table}

As was noted in \citet{Reese2008a}, the ratio $\Delta_{\tilde{\l}} /
\Delta_{\tilde{n}}$ decreases for increasing rotation rates.  Using
$\Delta_{\l}/\Delta_{n} = 1/2$ in the non-rotating case \citep{Tassoul1980}, and
the relationships between $\Delta_{\tilde{n}}$, $\Delta_{\tilde{\l}}$ and
$\Delta_n$, $\Delta_{\l}$ given in \citet{Reese2008a}, one finds a theoretical
value of $2$ for $\Delta_{\tilde{\l}}/\Delta_{\tilde{n}}$ when $\Omega = 0$.
Since the values in Table~\ref{tab:asymptotic} are much smaller, the ``small''
frequency separation is no longer small but comparable with the large frequency
separation, as was also observed in \citet{Lignieres2006} and
\citet{Lovekin2009}.

In the last column, the standard deviation between the asymptotic and numerical
frequencies is given.  It is defined as follows:
\begin{equation}
\left< \delta \omega^2 \right>^{1/2} = \sqrt{ \frac{1}{N_{\mathrm{modes}}}
\sum_{i=1}^{N_{\mathrm{modes}}} \left( \omega_i -\omega_i^{\mathrm{asymp.}}
\right)^2}
\end{equation}
where $\omega_i^{\mathrm{asymp.}}$ are the frequencies given by the
asymptotic formula and $\omega_i$ the numerical frequencies.  Although the
asymptotic formula captures the basic structure of the frequency spectrum (at
least for low values of $m$), there are differences which are larger than
observational error bars.  The main causes seem to be deviations resulting from
avoided crossings and also a slight variation of the azimuthal dependence of the
frequencies with $\tilde{\l}$ and $\tilde{n}$ (see following section).

\subsection{High azimuthal orders}

\begin{figure}[htbp]
  \includegraphics[width=0.49\textwidth]{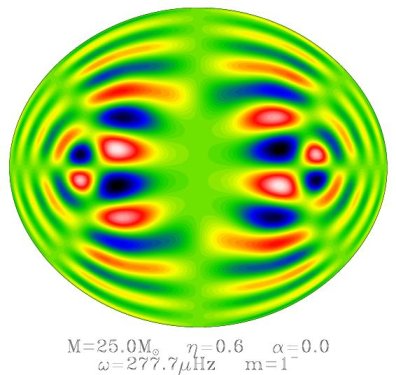} \hfill
  \includegraphics[width=0.49\textwidth]{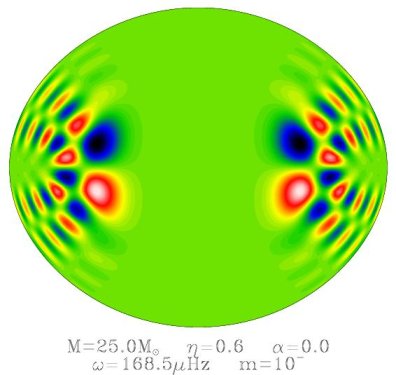}
  \caption{Two island pulsation modes, one with a low $m$ value (left)
           and the other with a high $m$ value (right).  Although the
           basic structure remains the same, the mode with a high azimuthal
           order is concentrated much closer to the equator.  This is analogous
           to what happens with sectoral modes in non-rotating stars when
           $m$ increases.}
  \label{fig:high_m}
\end{figure}

The results presented so far were based on pulsation modes with azimuthal orders
between -2 and 2.  However, island modes also exist for high values of $m$ as is
illustrated in Fig.~\ref{fig:high_m}.  As can be seen in the figure, high $m$
island modes have an analogous structure to their low $m$ counterparts except
that they are much closer to the equator.  This is similar to the behaviour of
sectoral modes in non-rotating stars.

\begin{figure}[htbp]
  \includegraphics[width=0.49\textwidth]{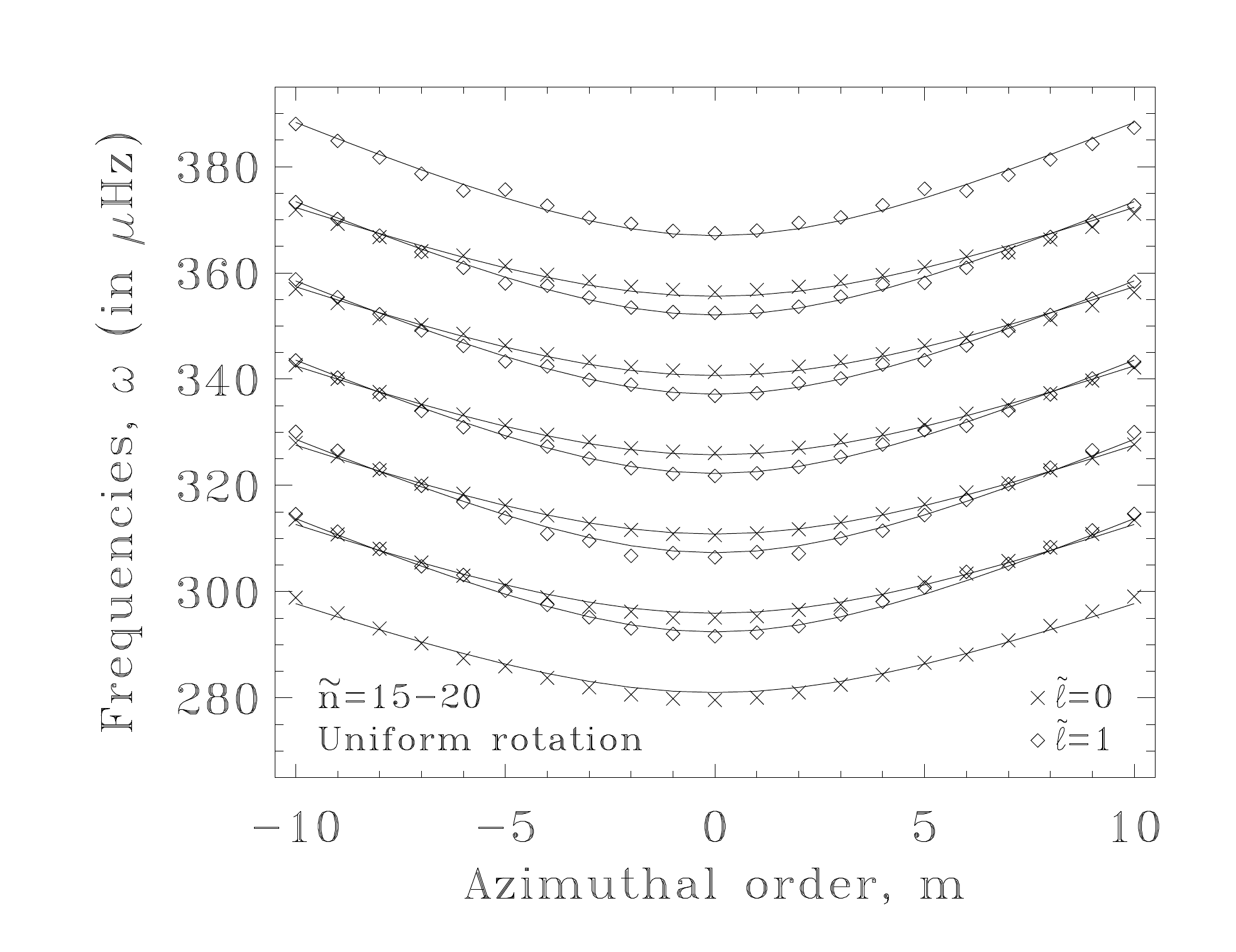} \hfill
  \includegraphics[width=0.49\textwidth]{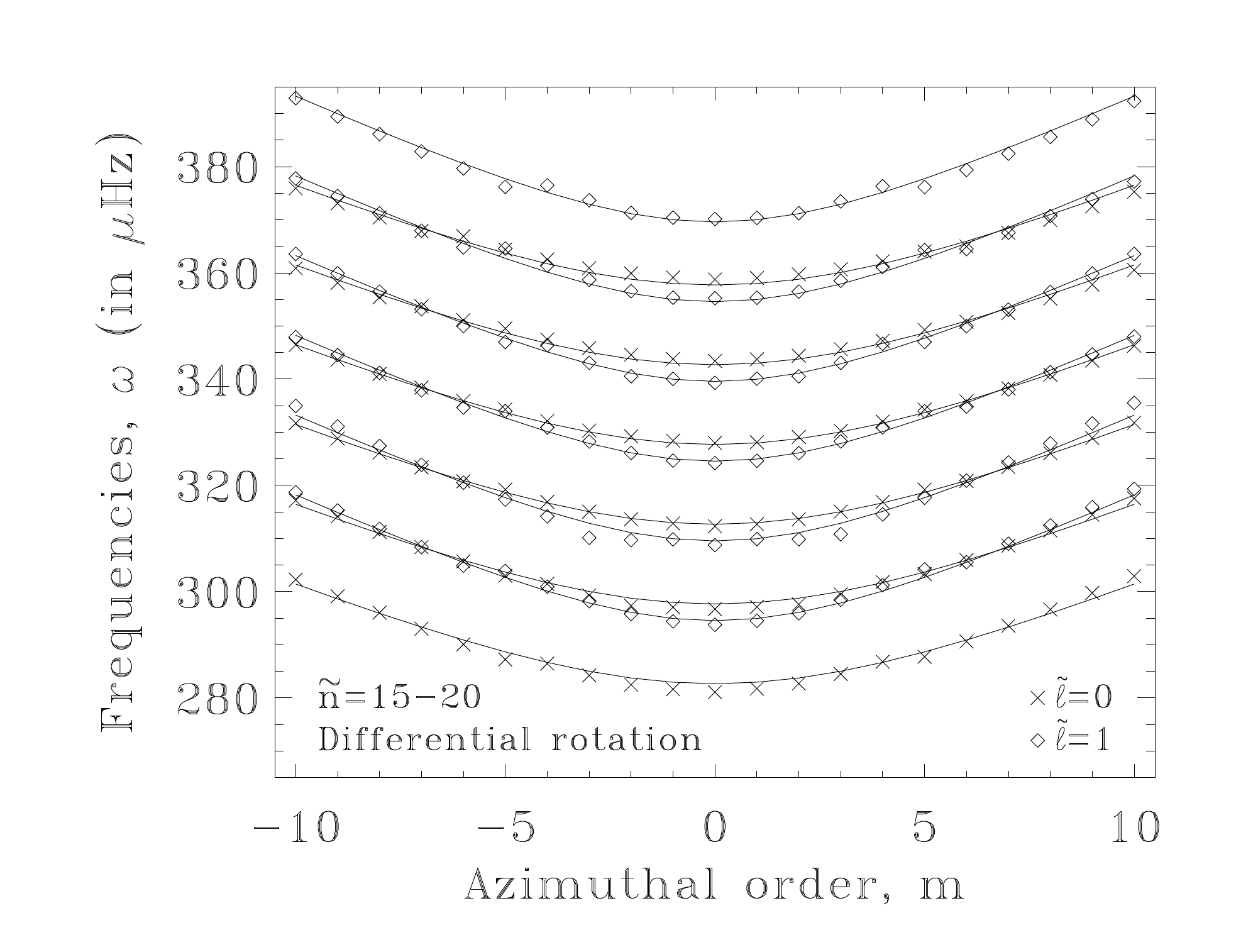}
  \caption{Pulsation frequency spectrum in a uniformly (left panel) and
  differentially (right panel) rotating model.  The radial order $\tilde{n}$
  goes from $15$ (bottom) to 20 (top), and $\tilde{\l} = 0$ and $1$. 
  Each hyperbola corresponds to a distinct pair
  $(\tilde{n},\tilde{\l})$.  The symbols represent the numerically calculated
  frequencies and the continuous lines correspond to a least-squares fit based
  on Eq.~(\ref{eq:asymptotic}).  Some of the irregular features in the numerical
  frequencies are caused by avoided crossings.}
  \label{fig:frequency_spectrum}
\end{figure}

Figure~\ref{fig:frequency_spectrum} shows two pulsation frequency spectra with
$\tilde{n}=15$ to $20$, $\tilde{\l} = 0$ to $1$ and $m=-10$ to $10$.  The left
plot is for a uniformly rotating model and the right one corresponds to
differential rotation.  The symbols represent the numerical frequencies and the
continuous lines are a least-squares fit based on the following formula:
\begin{equation}
  \omega_{\tilde{n},\,\tilde{\l},\,m} = 
               \tilde{n} \Delta_{\tilde{n}}
             + D_{\tilde{m}}(\tilde{\l}) \sqrt{m^2 + \mu(\tilde{\l})^2}
             - m \Omega_{\mathrm{fit}}
             + \tilde{\alpha}(\tilde{\l}).
\label{eq:asymptotic2}
\end{equation}
The term $m\Omega_{\mathrm{fit}}$ has been removed from both the numerical
frequencies and the fit in Fig.~\ref{fig:frequency_spectrum} so as to bring out
their more subtle azimuthal dependence.  In Eq.~(\ref{eq:asymptotic2}), the
term with $\Delta_{\tilde{m}}$ has been replaced so as to give the
frequencies a hyperbolic dependence on $m$, as is visually suggested by the
numerical frequencies.  Besides the modification to the azimuthal term,
the term $\Delta_{\tilde{\l}}$ has been removed but is compensated for by
allowing the parameters $D_{\tilde{m}}$, $\mu$ and $\tilde{\alpha}$ to depend on
$\l$. 

Table~\ref{tab:asymptotic2} gives the values of the different parameters used to
fit the frequency spectra in Fig.~\ref{fig:frequency_spectrum}. Comparing these
values with those in Table~\ref{tab:asymptotic} shows reasonable agreement,
provided one compares $\Delta_{\tilde{m}}$ and $\Delta_{\tilde{\l}}$ with
$D_m/2\mu$ and $\alpha(\tilde{\l}=1)-\alpha(\tilde{\l}=0)$, respectively,
where the expressions $D_m/2\mu$ comes from a Taylor expansion of
Eq.~(\ref{eq:asymptotic2}) around $m = 0$. Although the average deviations are
larger in this table, Eq.~\ref{eq:asymptotic2} is a better fit to the
frequencies than Eq.~\ref{eq:asymptotic}.  The reason for this apparent
contradiction is because the frequency sets used in Table~\ref{tab:asymptotic2}
cover a larger range of $m$ values.  Applying Eq.~\ref{eq:asymptotic} to these
expanded sets would yield $\left< \delta \omega^2
\right>^{1/2}/\Delta_{\tilde{n}} = 0.079$ and $0.088$ for the uniformly and
differentially rotating models, respectively.

An important difference between the two formulae, is that contrary to
what is suggested by Eq.~(\ref{eq:asymptotic}), the azimuthal dependence is
different for $\tilde{\l}=0$ and $\tilde{\l}=1$ modes.  A likely cause is the
fact that pulsation modes are closer to the equator at high $m$ values.  This
would then modify the path which intervenes in the time integrals used to
calculate $\Delta_{\tilde{\l}}$, when working with ray dynamics
\citep{Lignieres2008}.  As a result, a physically more relevant formula for the
frequencies would include a $\Delta_{\tilde{\l}}$ term which depends on $m$
rather than an azimuthal term which depends on $\tilde{\l}$.  Of course, a
quantitative calculation based on ray dynamics is needed to support this
explanation.

\begin{table}[htbp]
\begin{center}
\caption{Parameters for the asymptotic formula Eq.~(\ref{eq:asymptotic2}).}
\label{tab:asymptotic2}
\begin{tabular}{*{15}c}
\hline
\hline
\multicolumn{3}{c}{Stellar parameters} & & & &
\multicolumn{4}{c}{\dotfill $\tilde{\l}=0$ \dotfill} &
\multicolumn{4}{c}{\dotfill $\tilde{\l}=1$ \dotfill} & \\
\hline
$\displaystyle \frac{M}{M_{\odot}}$ & 
$\eta$ &
$\alpha$ &
$\!\!\!\begin{array}{c} \displaystyle \Delta_{\tilde{n}} \\ (\mu\mathrm{Hz}) \end{array}\!\!\!$ & 
$\displaystyle \frac{\Omega_{\mathrm{fit}}}{\Delta_{\tilde{n}}}$  &
$\displaystyle \frac{\Omega_{\mathrm{real}}}{\Delta_{\tilde{n}}}$  &
$\displaystyle \frac{D_{\tilde{m}}}{\Delta_{\tilde{n}}}$ &
$\displaystyle \mu$ &
$\displaystyle \frac{\tilde{\alpha}}{\Delta_{\tilde{n}}}$ &
$\displaystyle \frac{D_{\tilde{m}}}{2 \mu \Delta_{\tilde{n}}}$ &
$\displaystyle \frac{D_{\tilde{m}}}{\Delta_{\tilde{n}}}$ &
$\displaystyle \mu$ &
$\displaystyle \frac{\tilde{\alpha}}{\Delta_{\tilde{n}}}$ &
$\displaystyle \frac{D_{\tilde{m}}}{2 \mu \Delta_{\tilde{n}}}$ &
$\displaystyle \frac{\left< \delta \omega^2 \right>^{1/2}}{\Delta_{\tilde{n}}}$ \\
\hline
25 & 0.6 & 0.0 & 14.91 & 0.980 & 0.989       & 0.197 & 5.95 & 2.67 & 0.0166 & 0.234 & 5.11 & 3.41 & 0.0229 & 0.045 \\
25 & 0.6 & 0.2 & 15.01 & 0.962 & 0.953-0.993 & 0.221 & 6.04 & 2.50 & 0.0183 & 0.244 & 4.52 & 3.52 & 0.0270 & 0.052 \\
\hline
\end{tabular}
\end{center}
\begin{list}{}{}
\item[]  Values of the parameters from Eq.~(\ref{eq:asymptotic2}) used to fit
         the frequencies in Fig.~\ref{fig:frequency_spectrum}
         (\textit{i.e.} the ranges on the quantum numbers are
         $15\leq\tilde{n}\leq 20$, $0\leq\tilde{\l}\leq1$ and $-10\leq m \leq
         10$). These values are similar to those in Table~\ref{tab:asymptotic}
         (see text for details).  The two values of
         $\Omega_{\mathrm{real}}$ for the differentially rotating model (second
         line) are the lower and upper on the angular velocity, \textit{i.e.}
         the equatorial and polar rotation rates, respectively. The parameter
         $\Omega_{\mathrm{fit}}$ is twice as close to $\Omega_{\mathrm{real}}$
         as in Table~\ref{tab:asymptotic} for the uniformly rotating model
         ($\alpha = 0$) due to the inclusion of higher azimuthal orders.
\end{list}
\end{table}

\begin{figure}[htbp]
  \begin{minipage}{0.45\textwidth}
  \includegraphics[width=0.99\textwidth]{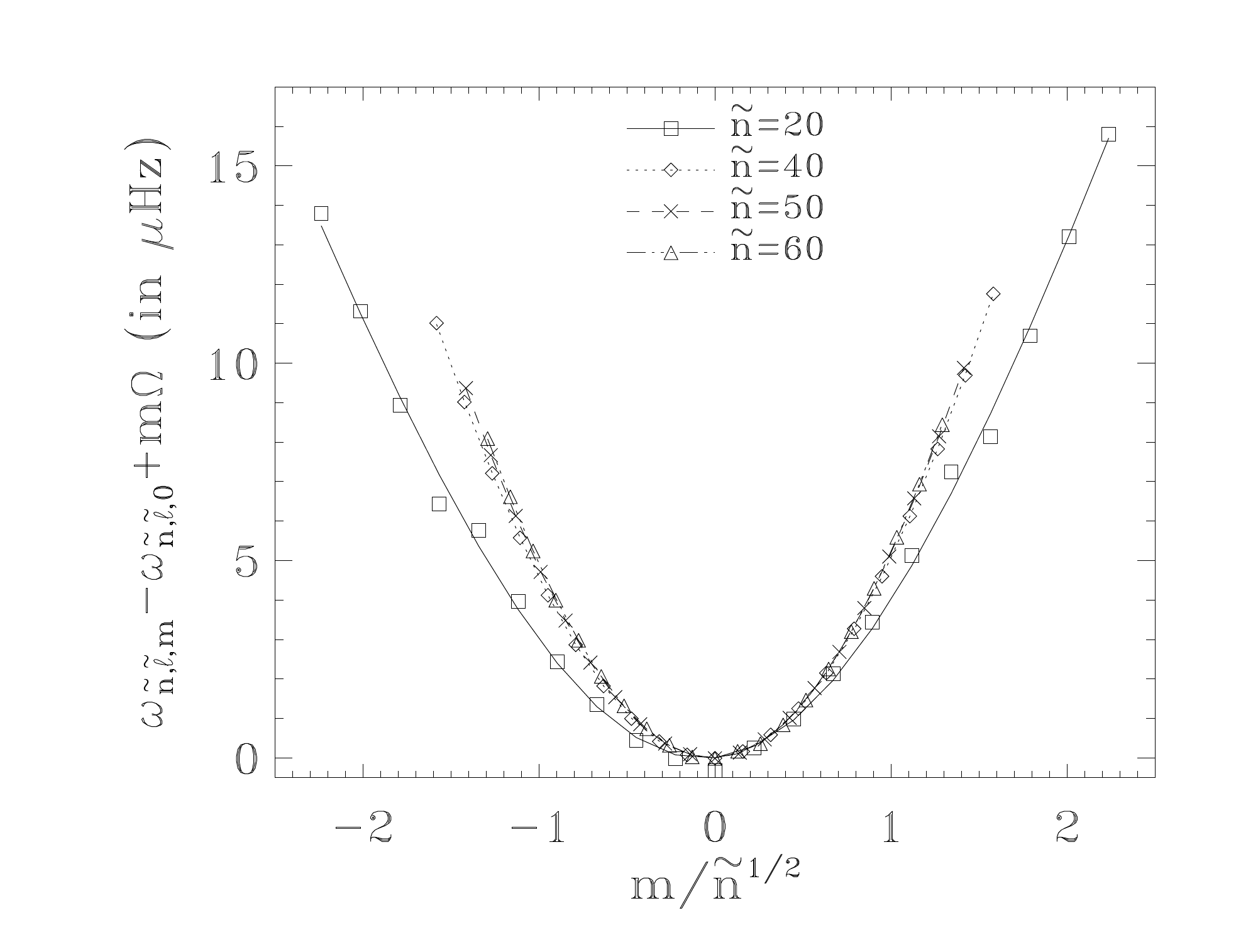}
  \end{minipage} \hfill
  \begin{minipage}{0.52\textwidth}
  \caption{Four sets of pulsation frequencies in a co-rotating frame in
  which $\tilde{n} = 20$, $40$, $50$ and $60$.  The other quantum numbers are
  $\tilde{\l}=0$ and $m=-10$ to $10$ for all sets.  The azimuthal order has been
  scaled by $1/\sqrt{\tilde{n}}$ as this causes the high order curves to overlap
  and reduces the difference between these curves and the $\tilde{n}=20$
  curve.}
  \label{fig:scaled_n}
  \end{minipage}
\end{figure}

It is also interesting to look at what happens when $\tilde{n}$ is increased to
a large value. Figure~\ref{fig:scaled_n} compares 4 sets of pulsation
frequencies corresponding to $\tilde{n} = 20$, $40$, $50$ and $60$. The symbols
represent the numerical frequencies and the continuous lines a fit based on
Eq.~(\ref{eq:asymptotic2}).  The frequencies are given in a co-rotating
frame and have been shifted so that the different curves are at $0$ for
$m/\sqrt{n} = 0$.  Plotting the frequencies as a function of
$m/\sqrt{\tilde{n}}$ rather than $m$ causes the curves associated with the high
order frequencies to overlap and reduces the difference between these curves and
the $\tilde{n} = 20$ curve.  These results suggest that as $\tilde{n}$ goes to
infinity, the azimuthal dependence of the co-rotating frequencies can be
described by a law of the form $n^{\delta}f\left(m/n^{\gamma}\right)$ where $f$
is a function and $2\gamma - \delta = 1$.

\subsection{An effective rotation rate}
\label{sect:omega_eff}

\begin{figure}[htbp]
  \begin{minipage}{0.45\textwidth}
  \includegraphics[width=0.99\textwidth]{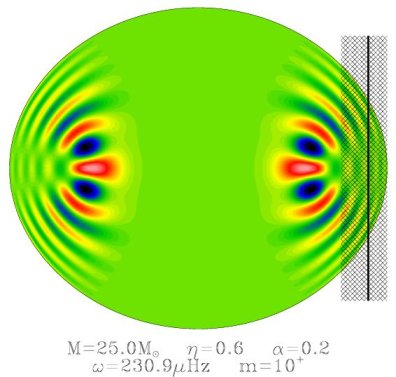}
  \end{minipage} \hfill
  \begin{minipage}{0.52\textwidth}
  \caption{The $(\tilde{n},\,\tilde{\l},\,m) = (20,0,10)$ island mode in
  the differentially rotating model.  The vertical thick line corresponds to the
  position where $\Omega_{\mathrm{fit}}$ is equal to the local rotation rate. 
  The hashed region on either side is an estimate of the error on this position,
  based on the difference between $\Omega_{\mathrm{fit}}$ and
  $\Omega_{\mathrm{real}}$ for the uniformly rotating model.}
  \label{fig:hashed_field}
  \end{minipage}
\end{figure}

Of particular interest is the parameter $\Omega_{\mathrm{fit}}$.  In the
uniformly rotating case, the term $-m\Omega_{\mathrm{fit}}$ represents, to first
order, the advection of the modes by stellar rotation.  The value given
for $\Omega_{\mathrm{fit}}$ in Table~\ref{tab:asymptotic2} is quite close to the
true rotation rate and only differs by $0.93 \%$, this difference probably
resulting from the Coriolis force.  In the differentially rotating case,
$-m\Omega_{\mathrm{fit}}$ can also be interpreted as an estimate of the
advection of the pulsation modes by stellar rotation.  The parameter
$\Omega_{\mathrm{fit}}$ would then be an average of the rotation rate in which
the weighting depends on the structure of the pulsation modes.  We will refer to
$\Omega_{\mathrm{fit}}$ as an effective rotation rate.  We can then use
Eq.~(\ref{eq:rotation}) to calculate the position $s_{\mathrm{fit}}$ where
$\Omega(s)$ is equal to $\Omega_{\mathrm{fit}}$ for the differentially rotating
model.  This is represented by the thick vertical line in
Fig.~\ref{fig:hashed_field} for the numerical value given in
Table~\ref{tab:asymptotic2}.  The hashed region on either side of this line is
an estimate of the error on this position using the difference between
$\Omega_{\mathrm{fit}}$ and $\Omega_{\mathrm{real}}$ from the uniformly rotating
model as a guide.

As can be seen from Fig.~\ref{fig:hashed_field}, $s_{\mathrm{fit}}$ is located
towards the outer regions of the star.  This means that $\Omega_{\mathrm{fit}}$,
when viewed as an average of the rotation profile, has a stronger weighting in
these outer regions.  This seems logical from the point of view of ray dynamics
because a sound wave travelling along a ray path will spend most of its time in
the outer regions of the star where the local sound velocity is lower.  As a
result, it will spend more time being advected by rotation in that region rather
than in an inner region.  Sound-travel times along ray paths have already been
used to establish an asymptotic expression for rotational kernels of high order
p-modes in spherical stars \citep{Gough1984}.

One of the best ways to confirm these ideas in a quantitative way is to
apply a variational formula which is valid for differential rotation.  Such a
formula has been established in \citet{Lynden-Bell1967}.  Here we give a
different and somewhat simpler expression which is only valid for conservative
rotation profiles:
\begin{eqnarray}
0 &=& \int_V (\omega_\mathrm{var} + m \Omega)^2 \rho_o \|\vect{\xi}\|^2 dV
+2 i \int_V (\omega_\mathrm{var} + m \Omega) \rho_o
\vect{\Omega} \cdot \left( \vect{\xi}^* \times \vect{\xi} \right) dV
\nonumber \\
 & & - \int_V \rho_o \left|\xi_s\right|^2 s \partial_s \left( \Omega^2 \right) dV
- \int_V \frac{|p|^2 dV}{\rho_o c_o^2}
- \int_V \rho_o N_o^2 \left|\xi_g\right|^2 dV 
+ \frac{1}{\Lambda}\int_{V_{\infty}} \| \grad \Psi \|^2 dV,
\label{eq:variational2}
\end{eqnarray}
where $\vect{\xi}$ is the Lagrangian displacement, $\xi_s$ the displacement 
component perpendicular to the rotation axis and $\xi_g$ the displacement
component in the same direction as the effective gravity.  In order to apply
this formula, it is necessary to calculate pulsation modes in terms of
$\vect{\xi}$ rather than $\vect{v}$, the Eulerian velocity perturbation.  This
has been done, and the relevant pulsation equations are described in
App.~\ref{sect:lagrange}.

\begin{figure}
  \begin{minipage}{0.45\textwidth}
  \includegraphics[width=0.99\textwidth]{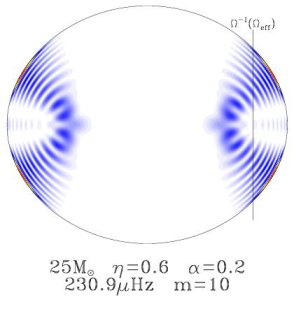}
  \end{minipage} \hfill
  \begin{minipage}{0.52\textwidth}
  \caption{A plot of the kernel associated with the
   $(\tilde{n},\,\tilde{\l},\,m) = (20,0,10)$ mode shown in
   Fig.~\ref{fig:hashed_field}.  As can be seen, the highest amplitudes are
   reached very near the surface, just like for p-modes in non-rotating stars. 
   The vertical line indicates the position of $s_\mathrm{fit}^\mathrm{eff} =
   \Omega^{-1}\left(\Omega_\mathrm{eff}\right)$ which can be compared with the
   position of $s_\mathrm{fit}$ as shown in Fig.~\ref{fig:hashed_field}.}
  \label{fig:kernel}
  \end{minipage}
\end{figure}

In the uniformly rotating case, the term that corresponds to the
advection of pulsation modes by rotation is $m\Omega$, which is contained in the
first integral.  By analogy, we can define an effective rotation rate for the
differentially rotating case as follows:
\begin{equation}
\int_V (\omega_\mathrm{var} + m \Omega)^2 \rho_o |\vect{\xi}|^2 dV
= (\omega_\mathrm{var} + m \Omega_\mathrm{eff})^2  \int_V \rho_o |\vect{\xi}|^2 dV
\end{equation}
Solving for $\Omega_\mathrm{eff}$ leads to the following expression:
\begin{equation}
\Omega_{\mathrm{eff}} = -\frac{\omega_\mathrm{var}}{m} + \frac{1}{m}
                         \sqrt{\int_V \left(\omega_\mathrm{var} 
                               + m \Omega\right)^2 \mathcal{K} dV},
\label{eq:Omega_eff}
\end{equation}
where
\begin{equation}
\mathcal{K} = \frac{\rho_o \|\xi\|^2}{\int_V \rho_o \|\xi\|^2 dV}.
\label{eq:kernel}
\end{equation}
When $m\Omega \ll 2\omega$, then $\Omega_{\mathrm{eff}}$ can be approximated by
$\int_V \Omega \mathcal{K} dV$.  This is similar to the first order perturbative
expression describing the advection of modes by slow rotation, except that the
kernel $\mathcal{K}$ has been defined from the eigenmode in the rotation star.
Figure~\ref{fig:kernel} shows a plot of the kernel associated with the mode in
Fig.~\ref{fig:hashed_field}.  As can be seen in the figure, the highest
amplitudes are reached very near the surface, just like for acoustic modes of
non-rotating stars.  Superimposed on the diagram is a vertical lines which
indicates the position of $s_\mathrm{fit}^\mathrm{eff} =
\Omega^{-1}\left(\Omega_\mathrm{eff}\right)$.  This can be compared with
$s_\mathrm{fit}$ which is plotted in Fig.~\ref{fig:hashed_field}.  As can be
seen from the two figures, $s_\mathrm{fit}^\mathrm{eff} < s_\mathrm{fit}$. This
difference comes from the fact that $s_\mathrm{fit}$ not only includes the
advection of modes by rotation but also the effects of the Coriolis force on the
mode frequencies.

\begin{figure}
  \begin{minipage}{0.45\textwidth}
  \includegraphics[width=0.99\textwidth]{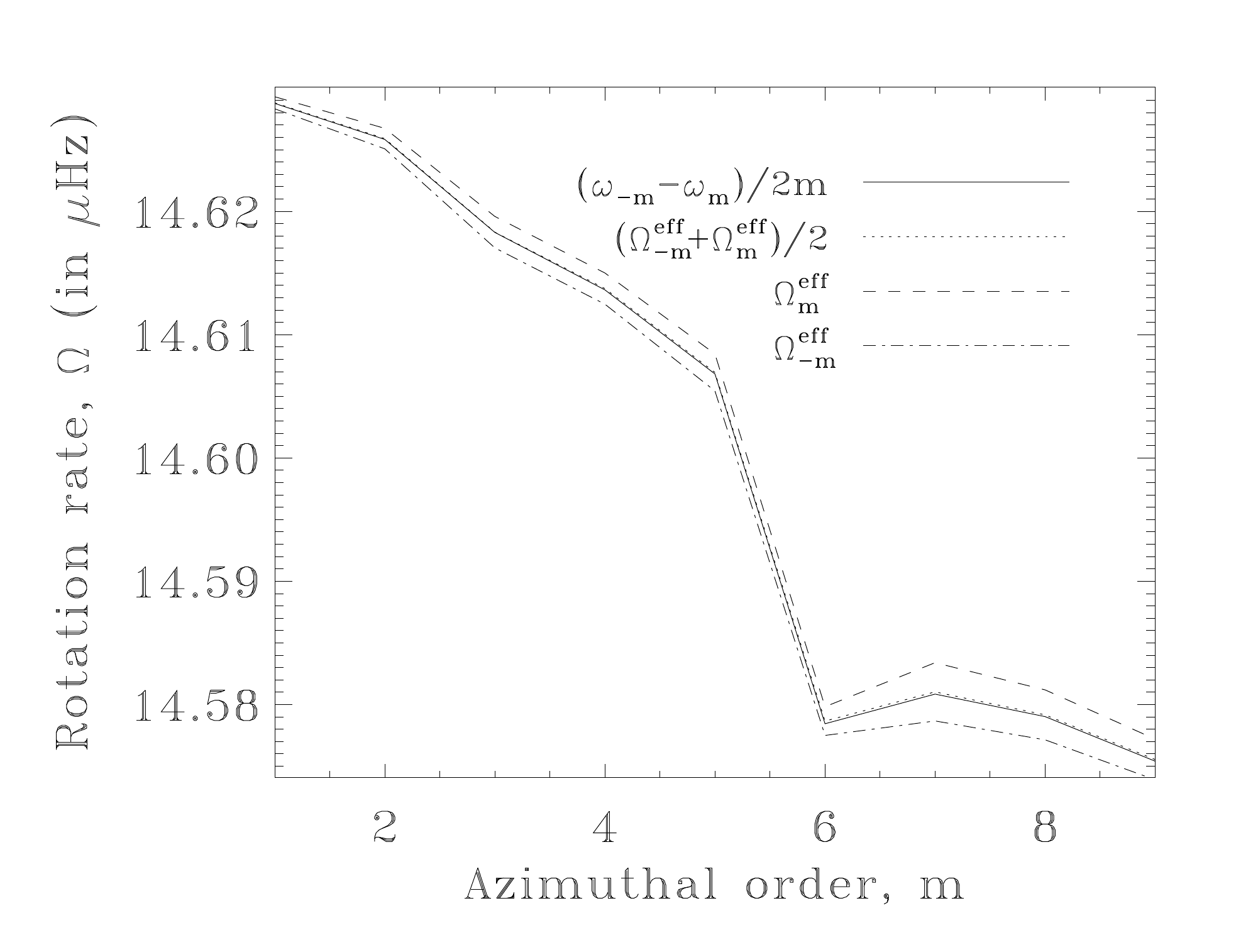}
  \end{minipage} \hfill
  \begin{minipage}{0.52\textwidth}
  \caption{Different measures of the effective rotation rate. The solid and
  dotted curves show an average over $m$ and $-m$ of $\Omega_{\mathrm{eff}}$
  based on the numerical frequencies and on Eq.~(\ref{eq:Omega_eff}).  The two
  last curves show the separate contributions from $m$ and $-m$, where $m$
  corresponds to retrograde modes and $-m$ to prograde modes.}
  \label{fig:Omega_eff}
  \end{minipage}
\end{figure}

The rotation rate $\Omega_{\mathrm{eff}}$ turns out to be a very good
indicator of the advection of modes by rotation.  This is illustrated in
Fig.~\ref{fig:Omega_eff}, which shows a comparison between  $\left(
\omega_{\tilde{n},\,\tilde{\l},\,m} - \omega_{\tilde{n},\,\tilde{\l},\,-m}
\right)/2m$ (solid line) and $\left( \Omega_{\mathrm{eff}}(m) +
\Omega_{\mathrm{eff}}(-m) \right)/2$ (dotted line), for a set of modes in which
the Coriolis force has been removed.  The relative difference between the two is
around $10^{-6}-10^{-5}$, making it difficult to distinguish  the two curves. 
The downward trend results from the fact that the pulsation modes are becoming
closer to the equator as $|m|$ increases.  The dent between $m=5$ and $m=6$ is
caused by an avoided crossing. Also, the curves corresponding to
$\Omega_{\mathrm{eff}}(m)$ and $\Omega_{\mathrm{eff}}(-m)$ have been included. 
The reason why these curves are not identical is because modes with azimuthal
orders $m$ and $-m$ are not identical even without the Coriolis force, because
of the differential rotation profile.

This naturally leads on to the idea of applying inversion theory to
probe the rotation profile using the rotational kernels defined in
Eq.~(\ref{eq:kernel}).  The quantity $\left( \omega_{\tilde{n},\,\tilde{\l},\,m}
- \omega_{\tilde{n},\,\tilde{\l},\,-m} \right)/2m$ is readily available from
observations, once an accurate mode identification has been done.  Furthermore,
it turns out that $\Omega^1_{\mathrm{eff}} = \int_V \Omega \mathcal{K} dV$ is a
very good approximation to $\Omega_{\mathrm{eff}}$, at least in the example
considered above, thereby allowing the use of linear inversion theory.  These
kernels will, nonetheless, need to refined so as to include the effects of the
Coriolis force.

\section{Highly differential rotation}
\label{sect:strongly_differential}

\begin{figure}
  \begin{center}
  \begin{tabular}{cc}
  \includegraphics[height=5cm]{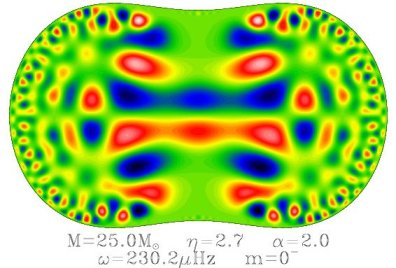} &
  \includegraphics[height=5cm]{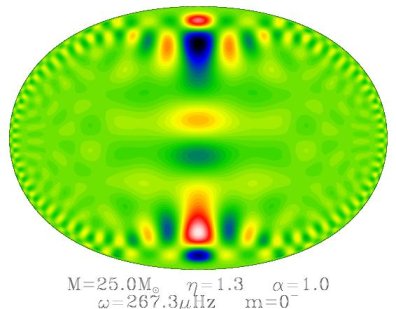}
  \end{tabular}
  \end{center}
  \caption{The left figure corresponds to a chaotic mode and the right one
  to what appears to be a whispering gallery mode in SCF models
  with a highly differential rotation profile.  No island modes are shown
  as they seem to have disappeared in the models (for low $m$).}
  \label{fig:highly_differential}
\end{figure}

When the rotation profile becomes highly differential, the stellar structure
becomes more and more deformed and the polar regions can, in some cases, become
concave.  These polar concavities result from the particular choice of rotation
profile as expressed in Eq.~(\ref{eq:rotation}) since they do not appear in
models where the rotation rate increases with distance from the rotation axis. 
This deformation naturally affects the structure and organisation of pulsation
modes.  Figure~\ref{fig:highly_differential} shows a chaotic and what appears
to be a whispering gallery mode in models where the rotation profile is very
differential.  No island modes are shown as they seem to have disappeared.  In
the more distorted configurations, even whispering gallery modes become
difficult to find. Instead, most of the modes are of a very chaotic nature.

\begin{figure}
  \begin{center}  \begin{tabular}{cc}
  \includegraphics[height=5cm]{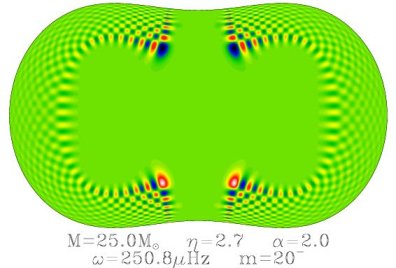} &
  \includegraphics[height=5cm]{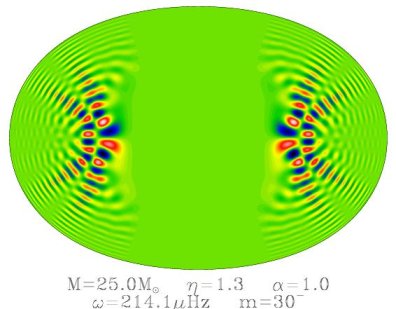}
  \end{tabular}
  \end{center}
  \caption{The two figures corresponds to pulsation modes with a high azimuthal
  orders in models with a highly differential profile.  The left figure
  corresponds to a whispering gallery mode and the right one to an island mode. 
  As can be seen in these plots, pulsation modes become less chaotic with
  increasing azimuthal order.}
  \label{fig:high_m_high_alpha}
\end{figure}

One way to counteract the effects of stellar distortion is to increase the
azimuthal order $m$.  Indeed, increasing the azimuthal order causes the
pulsation modes to become closer to the equator and move away from the poles
where stellar deformation is strongest.  As can be seen in
Fig.~\ref{fig:high_m_high_alpha}, highly regular whispering gallery modes exist
even in the most deformed configurations.  Also, for models with less
distortion, it is possible to find some island modes.  Nonetheless, such modes
are not likely to be visible in stars due to disk averaging effects.  Therefore,
if stars reach this degree of distortion, it will be very challenging to
interpret their pulsation spectra.

\section{Conclusion}
\label{sect:conclusion}

As has been shown in this paper, results concerning pulsation modes in rapidly
rotating polytropic models can be generalised to more realistic models based on
the Self-Consistent Field method \citep{Jackson2005, MacGregor2007} provided the
rotation profile is not too differential.  In particular, pulsation modes fall
into different categories, island, chaotic and whispering gallery
modes, each with their own characteristic geometry, in full agreement with
previous calculations based on ray dynamics \citep{Lignieres2008,
Lignieres2009}.  The frequencies of the island modes obey the same type of
asymptotic formula as those in polytropic models although a more careful
investigation of their $m$-dependence reveals a more complex behaviour than was
previously established.  This type of formula potentially provides a promising
way of identifying pulsation modes in rapidly rotating stars, especially at high
radial orders where the agreement between formula and frequency is very good
\citep{Reese2009b}.  Of course, when applying this formula to
observations, one should restrict themselves to modes with low $\tilde{\l}$ and
$m$ values, because cancellation effects reduce the visibility of modes with
more nodes on the surface.  As a result, the approximate form given by
Eq.~\ref{eq:asymptotic}, which is valid for low $m$ values, is sufficiently
accurate.

A useful by-product of the asymptotic formula is an estimate of the effective
rotation rate which gives the average advection of modes by rotation
when the rotation profile is mildly differential.  The obtained value indicates
a stronger weighting near the surface, where the local sound velocity is
smaller.  This goes hand in hand with the intuitive picture based on
ray dynamics that a sound wave is most advected in those regions where it spends
the most time.  A rigorous calculation based on a variational principle
yields rotation kernels which confirm this picture and help provide effective
rotation rates similar to the one obtained in the asymptotic formula, apart from
the effects of the Coriolis force on the frequencies.  These rotation kernels
could then be used in inversion methods to probe the rotation profile.

When the rotation profile is highly differential, pulsation modes tend to be
predominantly chaotic, probably as a result of the star's geometric distortion. 
Increasing the azimuthal order counteracts this effect by drawing the pulsation
modes closer to the equator thereby causing regular whispering gallery modes,
and in some cases, island modes, to reappear.  Nonetheless, such modes are not
likely to be visible due to disk averaging effects thus making pulsation spectra
in such stars difficult to interpret.

\section*{Acknowledgements}
The authors wish to thank the referee for valuable suggestions and
comments which helped to improve this article.
Many of the numerical calculations were carried out on the Altix 3700 of CALMIP
(``CALcul en MIdi-Pyr{\'e}n{\'e}es'') and on Iceberg (University of Sheffield),
both of which are gratefully acknowledged.  DRR gratefully acknowledges support
from the UK Science and Technology Facilities Council through grant
ST/F501796/1, and from the European Helio- and Asteroseismology Network (HELAS),
a major international collaboration funded by the European Commission's Sixth
Framework Programme.  The National Center for Atmospheric Research is a
federally funded research and development center sponsored by the U.S.~National
Science Foundation.

\bibliographystyle{aa}
\bibliography{biblio}

\appendix

\section{Pulsation equations based on the Lagrangian displacement}
\label{sect:lagrange}

In order to derive the Euler's equation in terms of the Lagrangian displacement,
we begin with Eq.~13 of \citet{Lynden-Bell1967} and calculate its Eulerian
perturbation in an inertial frame:
\begin{equation}
\rho_o \frac{D_o^2 \vect{\xi}}{Dt^2} - \rho_o \vect{\xi} \cdot \grad \left( \vect{v}_o
\cdot \grad \vect{v}_o \right) = -\grad p + \rho \vect{g}_\mathrm{eff} - \rho_o
\grad \Psi,
\label{eq:Lagrange}
\end{equation}
where $\xi$ is the Lagrangian displacement, and other quantities have the same
definitions as before.  The time derivation operator is defined as follows:
\begin{equation}
\frac{D_o \vect{\xi}}{Dt} = \dpart{\vect{\xi}}{t} + \vect{v}_o \cdot \grad \vect{\xi}
                          = \lambda \vect{\xi} + \vect{\Omega} \times \vect{\xi}
                            + im \Omega \vect{\xi}.
\end{equation}
Simplifying Eq.~\ref{eq:Lagrange} yields:
\begin{equation}
\valp^2 \rho_o \vect{\xi} + 2\valp \rho_o \vect{\Omega} \times \vect{\xi} + \rho_o \xi_s s \partial_s
\left( \Omega^2 \right) \vect{e}_s = -\grad p + \rho \vect{g}_\mathrm{eff} - \rho_o \grad
\Psi.
\label{eq:Lagrange2}
\end{equation}
This is the same equation as what is used in \citet{Lovekin2009}.  If
one uses the following relationship between $\vect{\xi}$ and $\vect{v}$
\citep[\eg][]{Christensen-Dalsgaard2003}:
\begin{equation}
\vect{v} = \dpart{\vect{\xi}}{t} + \vect{v}_o \cdot \grad \vect{\xi}
          -\vect{\xi} \cdot \grad \vect{v}_o,
\label{eq:v_xi_relation}
\end{equation}
it is possible to show that Eq.~\ref{eq:Euler2} and Eq.~\ref{eq:Lagrange2} are
equivalent.

In terms of the coordinate system described in Section~\ref{sect:geometry},
Euler's equation takes on the following explicit form:
\begin{eqnarray}
0 &=& \vlp^2 \rho_o \left[ \frac{\zeta^2 \rz \xiz}{r^2} 
        +\frac{\zeta \rt \xit }{r^2}\right]
        + 2i\vlp \frac{\Omega\zeta\sint}{r}\rho_o \xip \nonumber \\
  & &   - \rho_o s \left(\d_s \Omega^2\right) \rz \sint
          \left[ \frac{\zeta^2\sint}{r^2} \xiz
        + \frac{\zeta\left(\rt\sint+r\cost\right)}{r^2\rz} \xit \right]
        - \dz p + \frac{\dz P_o}{\rho_o} \rho
        - \rho_o \dz \Psi, \\
\noalign{\smallskip} 
\label{eq:spheroidal.Euler2_lagrange}
0 &=& \vlp^2 \rho_o \left[ \frac{\zeta^2 \rt \xiz}{r^2} + 
          \frac{\zeta(r^2+\rt^2) \xit}{r^2\rz} \right]
         +2i\vlp\frac{\Omega\zeta\left(\rt\sint+r\cost\right)}{r\rz}\rho_o\xip \nonumber \\
  & &    -\rho_o s \left(\d_s \Omega^2\right) \left(\rt\sint+r\cost\right)
          \left[ \frac{\zeta^2\sint}{r^2}\xiz +
          \frac{\zeta\left(\rt\sint+r\cost\right)}{r^2\rz}\xit \right]
         -\dt p + \frac{\dt P_o}{\rho_o} \rho
         -\rho_o \dt \Psi, \\
\noalign{\smallskip} 
\label{eq:spheroidal.Euler3_lagrange}
0 &=& \vlp^2 \rho_o \frac{\zeta}{\rz} \xip
         -2i\vlp\frac{\Omega\zeta^2\sint}{r}\rho_o \xiz
         -2i\vlp\frac{\Omega\zeta\left(\rt\sint+r\cost\right)}{r\rz}\rho_o\xit
         -\frac{\dphi p}{\sint}
         -\rho_o \frac{\dphi \Psi}{\sint}.
\end{eqnarray}
where $i\omega = \lambda$.  This is then supplemented by the the continuity equation,
\begin{equation}
0 = \rho + \div \left(\rho_o \vect{\xi} \right),
\end{equation}
the adiabatic relation,
\begin{equation}
0 = p + \vect{\xi}\cdot \grad p_o - c_o^2
\left(\rho + \vect{\xi} \cdot \grad \rho_o \right),
\end{equation}
and Poisson's equation,
\begin{equation}
0 = \lapl \Psi - \Lambda \rho,
\end{equation}
which in spheroidal geometry are:
\begin{eqnarray}
0 &=&   \rho
      + \frac{\zeta^2 \dz \rho_o \xiz
      + \zeta \dt \rho_o \xit}{r^2 \rz}
      + \frac{\zeta^2 \rho_o}{r^2 \rz}\left[
        \frac{\dz \left( \zeta^2 \xiz \right)}{\zeta^2}
      + \frac{\dt \left( \sint \xit \right)}{\zeta \sint}
      + \frac{\dphi \xip}{\zeta \sint}
       \right], \\
0 &=& \left( p - c_o^2 \rho \right)
    + \frac{\zeta^2}{r^2 \rz}
      \left( \dz p_o - c_o^2 \dz \rho_o \right) \xiz
    + \frac{\zeta}{r^2 \rz}
      \left( \dt p_o - c_o^2 \dt \rho_o \right) \xit, \\
0 &=& \frac{r^2 + \rt^2}{r^2 \rz^2}  \dzz \Psi
      +\cz  \dz \Psi
      -\frac{2\rt}{r^2 \rz} \dzt \Psi
      +\frac{1}{r^2} \lapl_{\theta \phi} \Psi
      - \Lambda \rho.
\end{eqnarray}
Using this set of equation yields the same modes and similar frequencies as
using the system of equations based on the Eulerian velocity perturbation
$\vect{v}$. This approach nonetheless has the advantage of yielding $\vect{\xi}$
which can be used more readily in the variational formula for differential
rotation.
\end{document}